\documentclass[prd,aps,twocolumn,floats,floatfix,nofootinbib,superscriptaddress] {revtex4-1}
\usepackage{graphicx}
\usepackage{dcolumn}
\usepackage{bm}
\usepackage{graphics}
\usepackage{slashed}
\usepackage{amssymb}
\usepackage{natbib}
\usepackage{amsmath}
\usepackage{url}
\usepackage[usenames,dvipsnames,svgnames,table]{xcolor}
\usepackage{color}
\usepackage{xcolor}
\usepackage{tabularx}
\usepackage{hyperref}
\usepackage{enumitem}
\usepackage{soul}
\usepackage{grffile}
\usepackage{algorithm}
\usepackage[noend]{algpseudocode}
\usepackage{mathtools}

\newcommand{\trK}{\mathrm{tr}K}
\newcommand{\trT}{\mathrm{tr}T}
\newcommand{\trKhat}{\mathrm{tr}{\hat K}}

\newcommand{\trAtilde}{\mathrm{tr}{\tilde A}}

\def\beq{\begin{equation}}
\def\eeq{\end{equation}}
\newcommand{\bea}{\begin{eqnarray}}
\newcommand{\eea}{\end{eqnarray}}
\newcommand{\mhduet}{{\texttt{MHDuet}~}} 
\newcommand{\mhduetX}{{\texttt{MHDuet}}} 
\newcommand{\FUKA}{{\texttt{FUKA}~}} 

\hypersetup{
	colorlinks=true,
	linkcolor=red,
	linkbordercolor=black,
	citecolor=blue,
	filecolor=magenta,      
	urlcolor=magenta,
}

\begin{document}

\title{Large Eddy Simulations of Magnetized Mergers of Black Holes and Neutron Stars}
\author{Manuel R. Izquierdo}
\affiliation{Departament  de  F\'{\i}sica $\&$ IAC3,  Universitat  de  les  Illes  Balears,  Palma  de  Mallorca,  Baleares  E-07122,  Spain}
\author{Miguel Bezares}
\address{Nottingham Centre of Gravity,
	Nottingham NG7 2RD, United Kingdom}
\address{School of Mathematical Sciences, University of Nottingham,
	University Park, Nottingham NG7 2RD, United Kingdom}
\author{Steven Liebling}
\affiliation{Long Island University, Brookville, New York 11548, USA}	
\author{$\text{Carlos Palenzuela}^1$}

\begin{abstract}
The LIGO-Virgo-Kagra collaboration has observed gravitational waves consistent with the mergers of a black hole and a neutron star, namely GW200105 and GW200115, providing evidence for such cataclysmic events. Although no electromagnetic counterpart was reported for either of these two events, under certain conditions 
black hole--neutron star mergers are expected to form a significant accretion disk and to produce both a short gamma ray burst and a kilonova, much as observed in the binary neutron star merger GW170817. 
Here, we extend our publicly available code \mhduet to study numerically the merger of a magnetized neutron star with a black hole. \mhduet
employs Large Eddy Simulation (LES) techniques to help capture the magnetic field amplification resulting from turbulence and other sub-grid scale dynamics in the post-merger stage. 
In particular, we simulate a merger with parameters favorable to producing an accretion disk, focusing on the formation and dynamics of the turbulent disk and the resulting magnetic field amplification. Following the tidal disruption and during the formation of the accretion disk, the magnetic field undergoes significant amplification driven by the Kelvin-Helmholtz instability, reaching strengths of more than $10^{14}\,\rm{G}$ from a realistic initial strength of $10^{11}\,\rm{G}$ in short timescales of approximately $20\,\rm{ms}$. Despite employing LES techniques with a finest resolution of $120\,\text{m}$ that is among the highest in black hole-neutron star mergers, it is still insufficient to demonstrate convergence of the magnetic field growth. Although the effects of the LES are here rather modest, we expect them to be more significant at higher resolution, as observed in binary neutron star merger simulations.
\end{abstract}

\maketitle

\section{Introduction}

Merger events involving black holes and neutron stars (BHNS) are expected to be significant sources of gravitational waves, providing crucial insights into various astrophysical phenomena. The LIGO-Virgo-Kagra collaboration (LVK) has confirmed this expectation by potentially observing at least two events with a low false-alarm-rate, labeled as GW200105 and  GW200115~\cite{Abbott:2023a}. These detections mark significant progress in our understanding of BHNS mergers, distinct from both binary black hole~(BBH) and binary neutron star~(BNS) mergers. In addition, the next generation of gravitational wave~(GW) detectors promises an increased number of BHNS observations~\cite{Gupta:2023evt}.
These early observations indicate that BHNS binaries tend to be highly asymmetric, showing significant differences in mass between the black hole~(BH) and the neutron star~(NS). The presence of this disparity in the low-range of total binary masses sets them apart from both BBH and BNS mergers, suggesting that they might have originated from different progenitor stars and follow different paths to formation (see, e.g., \cite{Broekgaarden:2022, Gompertz:2022, Michaely:2022}). 

During BHNS mergers, the NS is either directly engulfed by the BH or undergoes tidal disruption due to the tidal forces exerted by the BH while the star is outside the innermost stable circular orbit (ISCO). When disruption occurs, it releases neutron-rich material, which can be categorized into two types: a portion that remains gravitationally bound, forming an accretion disk around the BH remnant, and an unbound component known as ``dynamical ejecta." 
Multimessenger observations of BHNS mergers offer great scientific potential
just as GW170817 delivered many insights with a BNS merger.
Nevertheless, the possibility of observing an electromagnetic signal from a BHNS appears much less likely than with a BNS merger.
The difficulty arises from the need for the NS to undergo disruption before falling into the BH (for recent reviews, see~\cite{Foucart:2020ats,Kyutoku:2021icp}). However, the presence of matter in these mergers presents a unique opportunity for a broader spectrum of observations, shedding light on the properties of matter under extreme densities. Furthermore, one could assume instead that a better theoretical understanding %
could increase the number and likelihood of future observations of BHNS mergers. The challenge of distinguishing BHNS mergers from BNS mergers also motivates further study of BHNS mergers.

Numerical simulations are a powerful tool for studying BHNS mergers. In this work, we employ our code, \mhduet\cite{mhduet_webpage}, to study a BHNS binary in which the NS is tidally disrupted and part of its matter forms an accretion disk with significant mass. We focus on the post-merger dynamics of the magnetic field by using Large Eddy Simulation (LES), a technique in which the general relativistic magnetohydrodynamic (GRMHD) evolution equations are modified by including new terms to account for the unresolved, subgrid-scale (SGS) dynamics~\cite{Zhiyin:2015}. We derive these new terms by considering the gradient SGS model~\cite{leonard75,muller02a,grete16,grete17phd} based on mathematical arguments with no a priori physical assumptions. By including these SGS terms in the equations, one can recover, at least partially, the effects induced by the unresolved SGS dynamics over the resolved scales. The combination of high-order numerical schemes, high-resolution, and LES with the gradient SGS model allows us to improve the accuracy of the magnetic amplification in a turbulent regime. 

With our LES-enabled \mhduet code, previously employed to analyze the post-merger magnetic field in BNS mergers (see, e.g., Refs.~\cite{vigano19b,carrasco19,vigano20,aguilera2020,Palenzuela:2021gdo,Aguilera-Miret:2021fre, Palenzuela:2022kqk, 2023PhRvD.108j3001A}), we find significant differences in the magnetic field dynamics between BHNS and BNS mergers. While the presence of two magnetized compact objects in BNS mergers induces a strong Kelvin-Helmholtz instability (KHI) (for an overview of turbulence in BNS mergers,  see e.g., a very recent review~\cite{Radice:2024gic}), which leads to significant amplification of the magnetic field, the scenario is different in BHNS mergers. Here, the absence of a magnetic field associated with the BH combined with its lack of a surface precludes the onset of such instability when the compact objects come into contact. As a result, the expectations regarding the growth and amplification of the magnetic field during a BHNS merger are less clear.

This study follows previous studies that explored the magnetic field dynamics in BHNS mergers (see, e.g., Refs.~\cite{Chawla:2010sw, Etienne:2011ea, Etienne:2015cea, Kiuchi:2015qua, Paschalidis:2014qra, Ruiz:2017due, 10.1093/mnras/stab1824}). However, to the best of our knowledge, this work represents the first evolution of a BHNS binary using an LES scheme aimed at realistically capturing small-scale effects. These effects are crucial to model accurately the transfer of kinetic energy to magnetic energy that drives a significant amplification of the magnetic field. %
In both BHNS and BNS mergers, previous studies have typically started simulations with unrealistically strong magnetic fields (i.e., $\gtrsim 10^{14}\,\text{G}$), despite the consensus that old neutron stars in a binary system present magnetic fields $|B| \leq 10^{11}\,\text{G}$~\cite{Tauris:2017}. In these direct numerical simulations, an exceptionally high numerical resolution is required to capture these small-scale effects. However, this demanding approach is currently unfeasible with the existing generation of supercomputers and numerical relativity codes. 

In this work, simulations of the BHNS merger are carried out using several numerical resolutions, focusing mainly on studying the magnetic field dynamics. The effect of neutrinos produced by the hot accretion disk is not included in our simulations, although on short timescales of tens of milliseconds their impact should be almost negligible. After the merger, the magnetic field strength grows quite rapidly mostly driven by the KHI arising at the contact interfaces of the one-arm spiral structure formed after the disruption of the NS. This instability induces turbulent dynamics at shear interfaces that amplifies the magnetic field. The rotation of the fluid along with the turbulent dynamics then spreads these regions of large magnetic field throughout the disk.
Within approximately $20\,\text{ms}$ after the merger, the magnetic field is amplified from an initial strength of $10^{11}\,\text{G}$ to more than $10^{14}\,\text{G}$. Although the remnant disk becomes roughly axisymmetric by the end of our simulations with an angular velocity that decreases with radius, the magneto-rotational instability~(MRI) is only expected to become relevant much later, once the magnetic field becomes sufficiently large scale.

Even with LES techniques, we do not observe convergence of the magnetic field growth or its saturation at spatial resolutions of $120\,\text{m}$ and $240\,\text{m}$. Although the effects of the LES are rather modest, we expect, based on the use of LES with even higher resolutions in the simulations of BNS mergers that did demonstrate such convergence~\cite{Palenzuela:2022kqk}, that higher resolution simulations (below $100\,\text{m}$) would converge.
	
This paper is organized as follows. In Sec.~\ref{sec:setup}, we briefly summarize the method and initial setup employed for the numerical simulations. Following this, in Sec.~\ref{sec:results}, we present our numerical results from the BHNS merger. Finally, we conclude with a discussion in Sec.~\ref{sec:discussion}. We follow the usual convention according to which indices  written as $\{a,b,c,d\}$ $(\{i,j, k\})$ refer to spacetime (spatial) coordinates and adopt a metric signature $(-,+,+,+)$.

\section{Setup}
\label{sec:setup}
Here, we provide an overview of the equations of motion, initial data, and numerical setup of this work. The numerical simulations carried out in this paper have been performed using our publicly available \mhduet code \cite{mhduet_webpage}. Previously, it has been utilized to study mergers of magnetized NSs employing LES techniques~\cite{aguilera2020,Palenzuela:2021gdo,Aguilera-Miret:2021fre, Palenzuela:2022kqk, 2023PhRvD.108j3001A} and phase transitions in the merging of NSs binaries~\cite{Palenzuela:2021gdo}. In this version, a new feature has been added: the ability to evolve accurate initial data provided by the \FUKA library~\cite{Papenfort:2021hod,GRANDCLEMENT20103334}. 
More details on the formulation of the Einstein equations can be found in Ref.~\cite{bezares17} and about the numerical methods in Refs.~\cite{simf3,liebling20}.
%

\subsection{Evolution formalism}

The spacetime geometry is described by the Einstein equations within the covariant conformal Z4 formulation~(CCZ4) (for further details, see Refs.~\cite{alic,bezpalen}). 
The Z4 formulation~\cite{Z41,Bona:2003qn,Z44}  extends the Einstein equations by introducing a new four-vector $Z_a$ which measures the deviation from Einstein's solutions, namely
\begin{eqnarray}
R_{ab} &+& \nabla_a Z_b + \nabla_b Z_a   = 
8\pi \, \left( T_{ab} - \frac{1}{2}g_{ab} \,\trT \right) \nonumber \\
&+& \kappa_{z}  \left(  n_a Z_b + n_b Z_a - g_{ab} n^c Z_c \right)~,
\label{Z4cov}
\end{eqnarray}
where $T_{ab}$ is the stress-energy tensor describing the matter content. 
Note the presence of damping terms, proportional to the parameter $\kappa_{z}$, which enforce the dynamical decay of the constraint violations associated with $Z_a$~\cite{gundlach} when $\kappa_{z} >0$. The presence of these terms is crucial for achieving stable and accurate simulations.

To express Eq.~\eqref{Z4cov} as an evolution system, one must split the spacetime tensors and equations into space and time components using a $3+1$ decomposition. The line element %
can be decomposed as
\begin{equation}
ds^2 = - \alpha^2 \, dt^2 + \gamma_{ij} \bigl( dx^i + \beta^i dt \bigr) \bigl( dx^j + \beta^j dt \bigr)~, 
\label{3+1decom}  
\end{equation}
where $\alpha$ is the lapse function, $\beta^{i}$ is the shift vector, and $\gamma_{ij}$ is the induced metric on each spatial foliation, denoted by $\Sigma_{t}$. In this foliation,  we define the normal to the hypersurfaces  $\Sigma_{t}$ as $n_{a}=(-\alpha,0)$ and the extrinsic curvature $K_{ij} \equiv  -\frac{1}{2}\mathcal{L}_{n}\gamma_{ij}$,  where $\mathcal{L}_{n}$ is the Lie derivative along  $n^{a}$. After rewriting the Einstein equations as a  time-dependent partial differential equation system, a conformal transformation can be applied as follows
\begin{eqnarray}
\tilde{\gamma}_{ij} &=& \chi\,\gamma_{ij}~,
\label{eq_gammaij}\\
\tilde{A}_{ij}      &=& \chi\left(K_{ij}-\frac{1}{3}\gamma_{ij} \trK \right)~,
\label{eq_Aij}
\end{eqnarray}  
where the conformal metric $\tilde{\gamma}_{ij}$ has unit determinant, $\chi$ is a positive function called the conformal factor, $\tilde{A}_{ij}$ is the conformal trace-free extrinsic curvature, and $\trK=\gamma^{ij}K_{ij}$.
For further convenience, we reformulate some of the evolved
quantities as
\begin{eqnarray}
	\trKhat &\equiv& \trK - 2\, \Theta~, \\
	{\hat \Gamma}^i &\equiv& {\tilde \Gamma}^i + 2 Z^{i}/\chi~,
\end{eqnarray}  
where $\Theta \equiv - n_{a} Z^{a}$. The final set of evolution
fields is $\{ \chi, {\tilde \gamma}_{ij}, \trKhat, {\tilde A}_{ij}, {\hat \Gamma}^i, \Theta  \}$, whose explicit evolution equations can be
found in Refs.~\cite{bezpalen,Palenzuela:2022kqk,simf3}. 

Notice that the conformal transformation given by Eqs.~(\ref{eq_gammaij}-\ref{eq_Aij}) introduce two new constraints, 
\begin{equation}
\tilde{\gamma} = \det (\tilde{\gamma_{ij}})= 1~,~~ \trAtilde  =\tilde{\gamma}^{ij}\tilde{A}_{ij}=0~,
\end{equation}
which have to be satisfied at every time step. Following this formalism, one can enforce them dynamically by including additional damping terms to the evolution equations proportional to $\kappa_c >0$. In this work, we set $\kappa_{c}=4/M_{0},$ where $M_{0}$ is the Arnowitt-Deser-Misner~(ADM) mass of the system at the initial time. 
Following  Ref.~\cite{PhysRevD.104.084010}, we explore reasonable values for the other damping parameter $\kappa_{z}$,
finding that the best results are obtained when $\kappa_{z} \lesssim 0.4/M_{0}$. 
Finally, in order to achieve stable evolution in the presence of black hole punctures~\cite{Alic:2013xsa}, the damping terms are modified by substituting $\kappa_{z} \rightarrow \kappa_{z}/\alpha$ and $\kappa_{c} \rightarrow \kappa_{c}/\alpha$.

These equations must be complemented
with coordinate (or gauge) conditions. Here we consider the Bona-Mass\'o family of slicing conditions for the lapse~\cite{BM}, combined with the Gamma-driver shift condition~\cite{alcub}, namely
\begin{eqnarray}
\partial_t \alpha & =& \beta^i \partial_i \alpha 
- 2 \,\alpha f(\alpha) \, {\hat K}~, 
\\ 
\partial_t \beta^i & =& \beta^j \partial_j \beta^i + {3\over4} g(\alpha)\, {\hat \Gamma}^i - \eta\,\left(\beta^i - \beta_0^i \right)~,
\label{system3}
\end{eqnarray}
where $\eta=2/M_{0}$ is a damping parameter for the shift and $f(\alpha), ~g(\alpha)$ are free functions. In our case, we have set $f(\alpha)=1$, $g(\alpha)=4/3.$ In some cases, the asymmetric ejection of mass and/or radiation of  gravitational waves may give a kick to the final remnant. In these cases, we use a small shift in the coordinates $\beta_0^i$ to set the center of mass back to the origin of our coordinate system~\cite{2022PhRvD.106b3008H}. We initialize the lapse $\alpha=1$ and $\beta^{i}=0$ in our simulations.

We describe a magnetized perfect fluid through the energy-momentum tensor 
\begin{eqnarray}\label{stress-energy-mperfectfluid}
T_{ab} &=& \left[ \rho (1 + \epsilon) + p \right] u_{a} u_{b} + p g_{ab}  \\
&+& {F_{a}}^{c} F_{b c}
- \frac{1}{4} g_{a b} ~ F^{c d} F_{c d}~, 
\nonumber
\end{eqnarray}
where $\rho$ is the rest-mass density of the fluid, $\epsilon$ its specific
internal energy, $p$ its pressure, $u^{a}$ the velocity four-
vector, and $F^{ab}$ is the Maxwell tensor of the electromagnetic field (i.e., where ${}^{*}\!F^{ab}$ is the Faraday tensor). The equations of motion are given by the conservation laws
\begin{eqnarray}
\nabla_a(\rho u^{a})=0~,~~~ \nabla_a \, T^{ab}=0~,~~~
\nabla_a \, {}^*\!F^{ab}=0~.
\end{eqnarray}

Now, we apply a $3+1$-decomposition to the four-velocity vector $u^{a}$  by  decomposing it into components parallel and orthogonal to the normal vector $n^{a}$
\begin{equation}
u^{a} = W(n^{a} + v^{a})~,
\end{equation}
where $W=-n_{a}u^{a}$ is the Lorentz factor and $v^{a}$ is the three-velocity vector, both of them measured by Eulerian observers. 

The GRMHD evolution equations   for a magnetized, non-viscous, and perfectly conducting fluid can be recast in flux-conservative form~\cite{palenzuela15}, namely 
\begin{equation}
\partial_{t} \textbf{u}
+\partial_{k}\mathcal{F}^{k}(\textbf{u}) = \mathcal{S}(\textbf{u})~,
\end{equation} 
which allows the use of high-resolution shock-capturing (HRSC) methods to deal with the inherent shocks appearing due to the non-linearities of the equations. The conservative variables $\textbf{u}=\left\lbrace \sqrt{\gamma}D, \sqrt{\gamma}S^i , \sqrt{\gamma}U, \sqrt{\gamma}B^i \right\rbrace$  are functions of the primitives variables $\left\lbrace\rho,\epsilon,v^{i},B^i\right\rbrace,$ where $B^i$  represents the magnetic field components. The primitive and the conserved variables are related by the non-linear relations
\begin{eqnarray}
D &=& \rho W ~, \\
S^i &=& (h W^2 + B^2 ) v^i - (B^k v_k) B^i ~,\\
U &=& h W^2 - p + B^2 - \frac{1}{2} \left[ (B^k v_k)^2 + \frac{B^2}{W^2} \right]  ~,
\end{eqnarray}
where $h=\rho(1+\epsilon)+p$ is the enthalpy. The pressure $p$, required to close the system of equations, is obtained from the equation of state (EoS) as detailed in Sec.~\ref{ID}. 
This set of GRMHD evolution equations is supplemented with additional fields.
A field $\phi$ is added to implement hyperbolic divergence cleaning that damps violations of the solenoidal constraint on the magnetic field~\citep{palenzuela18}.

In the context of LES techniques, the evolution equations are modified to account for the unresolved SGS dynamics. Within this approach, the GRMHD equations are modified by including new terms derived by using the gradient SGS model, which relies on the mathematical Taylor expansion of the non-linear fluxes of the magnetohydrodynamics~(MHD) equations as a function of the resolved fields. The inclusion of these SGS terms in the equations allows for the recovery, at least partially, of the effects that the unresolved sub-grid dynamics induce over the resolved scales, increasing the effective resolution of a numerical simulation~\cite{vigano20,aguilera2020}.

Following previous studies, and since we are mostly interested in the magnetic field dynamics, we here include only a description of the SGS term appearing in the induction equation, namely
\begin{eqnarray}
&& \partial_t (\sqrt{\gamma} {B}^i) + \partial_k [\sqrt{\gamma}(- \beta^k {{B}}^i  +  \beta^i {{B}}^k) 
\nonumber \\
&& \quad\quad\quad\quad + \alpha \sqrt{\gamma} ({\gamma}^{ki} {{\phi}} + {M}^{ki} - {\tau}^{ki}_{M} )] = \sqrt{\gamma} {{R_B}}^i ,
\label{evol_B_sgs}
\end{eqnarray}
where $M^{ki} = B^{i} v^{k} - B^{k} v^{i}$, while the source term ${{R_B}}^i$ is related to the divergence cleaning field (see Ref.~\cite{vigano20} for its explicit expression). The SGS tensor is given by %
\begin{equation}
\tau^{ki}_{M} = -~{\cal C_M}~\xi \, H_{M}^{ki} ~~, \label{eq:sgs_gradient}
\end{equation}
where again the explicit expressions for the tensor $H_{M}^{ki}$ in terms of field gradients are quite lengthy but can be found in the Appendix~A of Ref.~\cite{aguilera2020}.

Although the coefficient ${\cal C_M}$ is meant to be of order one for a numerical scheme having a mathematically ideal Gaussian filter kernel and neglecting higher-order corrections, the value that best mimics the feedback of small scales onto the large scales in a LES can differ depending partially on the numerical methods employed and on the specific problem, as discussed in Refs.~\cite{vigano19b,carrasco19,vigano20, aguilera2020}. In this work, we set ${\cal C_M}= 8$, which has been shown to reproduce the magnetic field amplification demonstrated by very high-resolution, non-LES simulations more faithfully for our numerical schemes~\cite{vigano20,aguilera2020}. Note that less dissipative numerical schemes would need smaller values of ${\cal C_M}$. Finally, the coefficient $\xi= \gamma^{1/3} \Delta^2/24$ is proportional to the spatial grid-spacing squared---typical for SGS models---and ensures convergence to the continuous limit by construction.

The conversion (or inversion) from the conserved fields to
the primitive ones, following previous works~\cite{aguilera2020,Aguilera-Miret:2021fre,Palenzuela:2021gdo,2023PhRvD.108j3001A}, is performed by employing the robust procedure introduced in Ref.~\cite{Kastaun:2020uxr}. Notice that failures in the inversion procedure might occur for different reasons, most commonly because of inaccuracies either in the fluid or in the spacetime metric fields. Most of these failures usually occur in very rarefied and/or highly magnetized mediums, or very close to the puncture, well inside the black hole horizon. Since the failures near the punctures are not handled automatically by the inversion procedure, we have implemented a smooth dissipation on the fluid fields inside the apparent horizon to alleviate such problem. This is well-motivated since this region is causally disconnected from the exterior. This modification in the interior of the BH is applied through a smooth kernel function defined as 
\begin{equation}
G(\alpha)=\exp\left[-\left(\frac{\alpha}{\alpha_{0}}\right)^{p}\right]~,
\end{equation}
where $\alpha_{0}$ sets the threshold where the new terms are relevant and $p$ indicates the sharpness of the transition. We commonly use $\alpha_{0}=0.15$ and $p=4$. For moderate spins $a/M \leq 0.5,$ stability is achieved by simultaneously damping and smoothing out the fluid fields inside the horizon by using the following driver terms
\begin{equation}
\partial_t \textbf{u}  = \rm{r.h.s}(\textbf{u}) - \alpha \sigma_\mathrm{BH} G(\alpha) 
\left[ \textbf{u}  + \Delta x\, D^2 U \right]~,
\label{rhs_g_1}
\end{equation}  
where $\rm{r.h.s}(\textbf{u})$ stands for the original right-hand-side of the  conservative variables, $\sigma_\mathrm{BH}$ is a damping parameter and $D^2$ is a second order discrete Laplacian operator.

More aggressive measures are required for larger spins $a/M > 0.5$,  since the metric $\gamma_{ij}$ might lose its positive-definiteness property during the evolution due to numerical inaccuracies~\cite{Etienne:2011ea,PhysRevLett.97.141101}.
In those cases, we enforce flatness of the conformal metric in the interior of the black hole, ensuring that the metric is always positive definite inside the horizon. This is achieved again by introducing a driver-term in the evolution equations for the conformal metric, namely 
\begin{equation}
\partial_t {\tilde \gamma}_{ij}  = \rm{r.h.s}({\tilde \gamma}_{ij})- \alpha\kappa_{\gamma} G(\alpha)(\tilde{\gamma}_{ij} - \eta_{ij})~,
\label{rhs_g_2}
\end{equation}  
where $\kappa_{\gamma}$ is a damping parameter and $\eta_{ij}$ is the flat metric in Cartesian coordinates. Since we are considering only moderate spins in this paper, we did not need to activate these last terms.

\subsection{Initial Data}\label{ID}

We build initial data consisting of a BH and a NS in a quasi-circular orbit using the public library \FUKA \cite{Papenfort:2021hod,GRANDCLEMENT20103334}. \FUKA is an initial data solver of the eXtended Conformal Thin-Sandwich formulation of Einstein's field equations for various compact object configurations, including spinning BHNS binaries. The neutron star is modeled employing a piece-wise polytropic EoS representing the cold part of the APR4 EoS.

Our goal is to study a binary that forms a massive accretion disk $M_{\rm{disk}} \approx 0.1\,M_{\odot}$ around the final black hole.
Early numerical simulations with non-spinning BHNS suggested~\cite{Shibata_2008} that tidal disruption occurs outside the ISCO for $q \geq 4$ with a stiff EoS (i.e., $R_\mathrm{NS}\approx 13\,\rm{km}$) or for $q \geq 3$ with a soft EoS (i.e., $R_\mathrm{NS}\approx 11\,\rm{km}$), where $q \equiv M_\mathrm{BH}/M_\mathrm{NS}$ is the mass-ratio of the binary. The effect of the black hole spin $a \equiv J/M$ was studied in Ref.~\cite{Kyutoku_2015}. In particular, Table~III of Ref.~\cite{Kyutoku_2015} shows that, with the soft EoS APR4 and $q=3$, the disk mass  is only $M_\mathrm{disk} \approx 4 \times 10^{-4}\,M_\odot$ for non-spinning black holes but increases significantly to $M_\mathrm{disk} \approx 0.08\,M_\odot$ when $a/M=0.5$.

Motivated by these results, we construct initial data corresponding to this latter setup. The mass of the black hole is $M_\mathrm{BH}=4.05\,M_{\odot}$ and its spin is $a/M=0.5$. The mass of the neutron star $M_\mathrm{NS}=1.35\, M_{\odot}$ and its areal radius is
$R_\mathrm{NS} = 11.2\,\rm{km}$. The binary is initially in a quasi-circular orbit with a separation of $44\,\rm{km}$. 
Half an orbit before the merger, we set an initial magnetic field that is  a purely poloidal dipolar field confined to the interior of the neutron star. The average magnetic field strength around the time of the disruption is $10^{11} \text{G}$, several orders of magnitude less than the large initial fields used in other simulations (e.g., \cite{Etienne:2011ea,Kiuchi:2015qua,Hayashi:2022cdq,2022PhRvD.106b3008H}) but within the observed upper range of the oldest known NSs.

\subsection{Numerical methods}

We use our code \mhduetX, generated automatically by the platform {\sc Simflowny} \cite{arbona13,arbona18} and based on the {\sc SAMRAI} infrastructure \cite{hornung02,gunney16}, which provides the parallelization and the adaptive mesh refinement. The code, which has been extensively tested in several scenarios \cite{palenzuela18,vigano19,vigano20,liebling20}, employs the following numerical methods: fourth-order-accurate operators for the spatial derivatives in the SGS terms and in the Einstein equations (the latter are supplemented with sixth-order Kreiss-Oliger dissipation); a HRSC method for the fluid, based on the  Lax-Friedrich flux splitting formula \cite{shu98} and the fifth-order reconstruction method MP5 \cite{suresh97}; a fourth-order Runge-Kutta scheme with sufficiently small time step $\Delta t \leq 0.4\,\Delta x$; and an efficient and accurate treatment of the refinement boundaries when sub-cycling in time~\cite{McCorquodale:2011,Mongwane:2015}.  A complete assessment of the implemented numerical methods can be found in Refs.~\cite{palenzuela18,vigano19}.

\begin{table}[t]
	\begin{tabular}{| c | c | c | c | c |}			
	\hline
	Case & ${\cal C_M}$	& ~~~$L^{x|y}$ [km]~~~& ~~~$L^{z}$ [km] &~~~ $\Delta^\mathrm{disk}_\mathrm{min}$ [m]
	\\ \hline
	{\tt LR} & 8 & [-90, 90] & [-90, 90] & 480 \\
	{\tt MR} & 8 & [-74, 74] & [-60, 60] & 240 \\
	{\tt HR} & 8 & [-52, 52] & [-22, 22] & 120 \\
	{\tt MRnL} & 0 & [-74, 74] & [-60, 60] & 240 \\
	\hline
    \end{tabular}
	\caption{{\em Simulation Parameters.} The simulations are evolved within a cubic coarse level spanning $\left[-1204\, \mathrm{km},1204\, \mathrm{km}\right]^3$ which has 7 additional levels of refinement.
        The different cases have a similar mesh refinement setup except for the sizes, but not the resolution, of the two finest levels.  The table shows the horizontal span of the finest level, $L^{x|y}$ and its
         vertical span, $L^z$. For all runs, the BHs are always contained in the finest level. However, because these two finest levels are different for
        the three cases, the bulk of the post-merger accretion disk is only 
         covered by this finest level for the highest resolution run. We therefore list
         the finest grid spacing, $\Delta^\mathrm{disk}_\mathrm{min}$, achieved within the bulk of the disk. 
         Each setup is adopted just before merger, while the inspiral phase is common to all of them. We have also included a run of medium resolution (\texttt{MRnL}) with no LES  for comparison purposes.}
	\label{tab:models}
\end{table}

The binary is evolved in a cubic domain of size $\left[-1204\, \mathrm{km},1204\, \mathrm{km}\right]^3$. 
During the inspiral, the binary is fully covered by the coarse grid, $6$ fixed mesh refinement~(FMR) and 1 adaptive mesh refinement~(AMR) levels,  reaching a minimum resolution $\Delta x_\mathrm{min}=120\, \rm{m}$ within the NS. Each consists of a cube with twice the resolution of the next larger one. At the merger time, we change the grid structure to achieve different resolutions in the disk, allowing us to study how the disk solution depends on the grid spacing.

All runs contain the black holes within the finest level; it is only the extent of
the finest two levels that differ.
In the high resolution (\texttt{HR}) case, %
the densest part of the remnant disk is also contained within the finest level with resolution $\Delta x_\mathrm{min}=120\, \rm{m}$. 
Note that this resolution is comparable to the highest resolution employed, to the best of our knowledge, in previous BHNS simulations~\cite{Kiuchi:2015qua}.
In the medium resolution (\texttt{MR}) case, %
the remnant disk is instead covered by the second finest level with resolution $\Delta x=240\, \mathrm{m}$. 
Finally, in the low resolution (\texttt{LR}) simulation the grid spacing in the disk is $\Delta x=480\, \mathrm{m}$. Notice that, during tidal disruption and disk formation, the  refinement grids remain fixed %
providing uniform resolution throughout the one-arm spiral and the shear layers. The specific values of these grid parameters, for the different resolutions considered here, can be found in Table~\ref{tab:models}. This setup allows us to study resolution effects
of the solution in the bulk of the remnant disk, which is our primary interest while conserving computational resources used for the outer envelope. 

\begin{figure*}[!ht]
	\includegraphics[width=18cm]{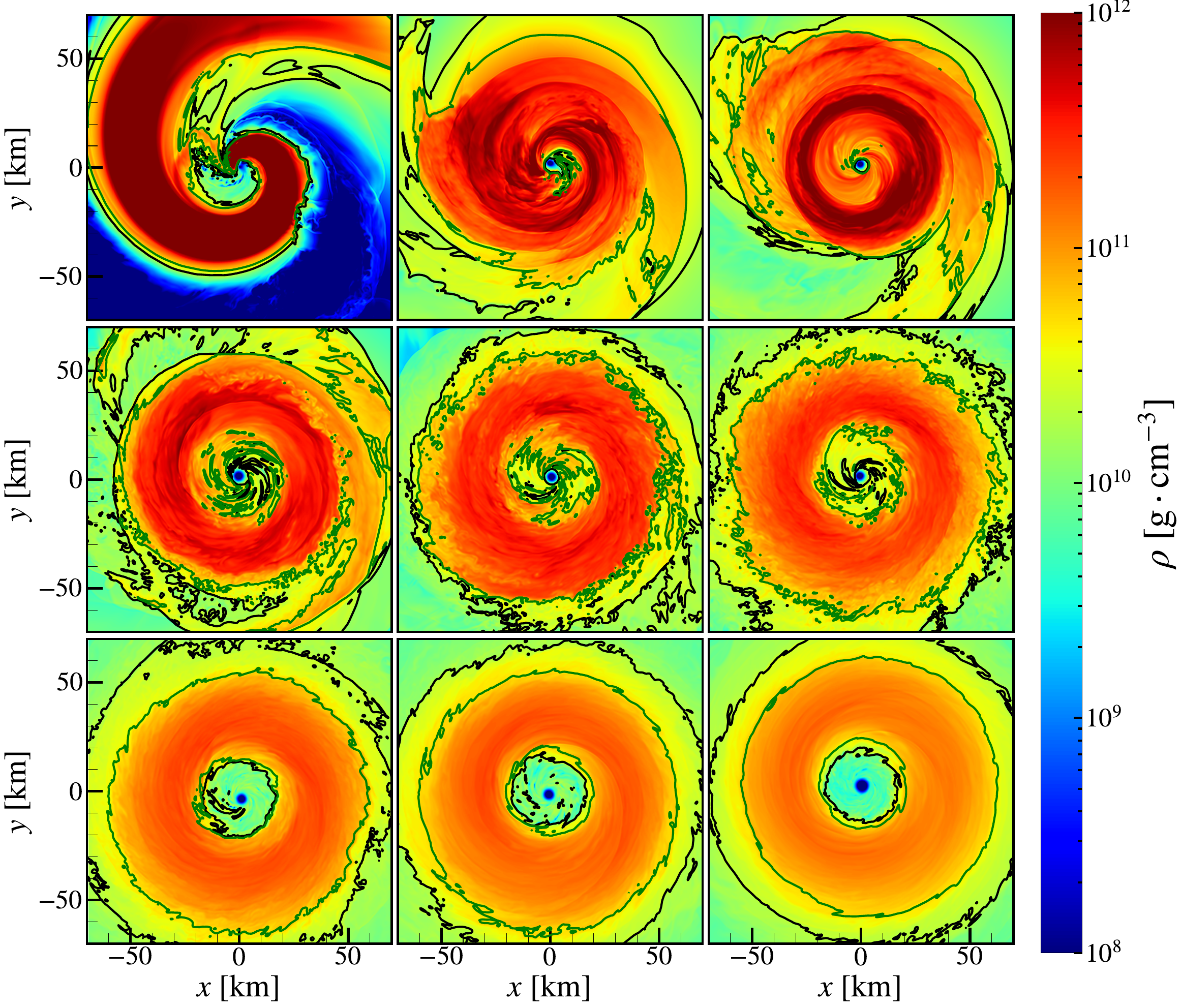}
	\caption{\emph{Evolution of the density in the orbital plane}. Snapshots of the rest-mass density at times $t = (0.5, 3, 4, 5.5, 7, 9, 16, 20, 25)$ ms (from left to right and top to bottom) with contour lines for
		$\rho = 2\times 10^{10} \text{g cm}^{-3}$ (black) and $\rho = 5\times 10^{10} \text{g cm}^{-3}$ (green).
		The neutron star is tidally disrupted, forming a one-arm spiral structure that wounds around the black hole. Shear layers are present when the head of this structure collides with its own tail and forms a turbulent and rapidly rotating torus, which relaxes to a roughly axi-symmetric disk by the end of our simulations.\label{RHOF}}
\end{figure*}
\subsection{Analysis quantities}

Here we summarize the analysis quantities that we use to monitor the dynamics in our simulations, described in more detail in Ref.~\cite{Palenzuela:2021gdo}. 

First, we compute global quantities, integrated over the whole computational domain, such as the total baryonic mass $E_{\rm{bar}}$, magnetic energy $E_{\rm{mag}}$, thermal energy $E_{\rm{th}}$, and rotational kinetic energy $E_{\rm{rot}}$. More physical information can be obtained by calculating suitable averages of certain quantities in different regions. In particular, we compute averages of the magnetic field magnitude, as well as its poloidal and toroidal components; the fluid angular velocity $\Omega \equiv \frac{d\phi}{dt} = \frac{u^{\phi}}{u^t};$ 
and the plasma beta parameter, $\beta \equiv \frac{2 p}{B^2}$ (not to be confused with the shift vector, $\beta^i$). The averages for a given quantity $\mathcal{Q}$ over a certain region ${\cal N}$ will be denoted generically by
\begin{eqnarray}
	\langle \mathcal{Q}\rangle_a = \frac{\int_{{\cal N}} \mathcal{Q} \, d{\cal N}}{\int_{{\cal N}} d{\cal N}}~,
\end{eqnarray}
where ${\cal N}$ stands for a volume $V$ or a surface $S$, 
and the integration is restricted to regions where the mass density is above $10^a\, \text{g cm}^{-3}$. Therefore, we define averages over the bulk or densest region of the remnant disk as $\langle \mathcal{Q} \rangle_{10}$ (actually, here we will abuse the notation and use it to indicate densities $\rho \geq 2 \times 10^{10}\, \text{g cm}^{-3}$), and averages over the whole disk (i.e. including the corona or envelope) as $\langle \mathcal{Q} \rangle_{8}$ (i.e., densities $\rho \geq 10^{8}\, \text{g cm}^{-3}$). 
Surface integrals are carried out over a cylinder $S$ with the axis passing through the centre of mass and orthogonal to the orbital plane. 

Further information can be obtained from the distribution of the kinetic and magnetic energies over the spatial scales (i.e., the kinetic and magnetic spectra), and defined respectively as:
\begin{eqnarray}
&& \mathcal{E}_k(k) = \frac{L^3 4\pi}{(2\pi)^3N^6} \langle k^2|\widehat{\sqrt{\rho}\vec{v}}|^2(\vec{k})\rangle_{k}~, \label{eq:spectra_k}\\
&& \mathcal{E}_m(k) = \frac{L^3 4\pi }{(2\pi)^3N^6}\langle k^2|\hat{\vec{B}}|^2(\vec{k})\rangle_{k}~, ~\label{eq:spectra_b}
\end{eqnarray}
where $L=140\, \text{km}$ is the size of a domain containing the most significant part of the disk remnant over which the fast Fourier transform is performed (indicated here by the wide hat), $k \in [2\pi/L,\pi/\Delta]$ is the radial wavenumber, $\langle~\rangle_k$ is the average over the spherical surface in the Fourier space corresponding to a fixed radial wavenumber $k_r = k$, and $N = L/\Delta$ is the number of points in each direction (ensuring the correct normalization). We also calculate the poloidal and toroidal contributions separately for the magnetic spectra. Further details of the numerical procedure to calculate the spectra can be found in~\cite{aguilera2020,vigano19,vigano20}.
With these spectrum distributions, we can define the spectra-weighted average wavenumber
\begin{equation}
	\langle k \rangle \equiv \frac{\int_k k\,{\cal E}(k) \,dk} {\int_k {\cal E}(k)\, dk}~,
\label{eq:coherence}
\end{equation}
with an associated length scale $\langle L\rangle = 2\pi/\langle k\rangle,$ which represents the typical coherence scale of the structures present in the field.

Finally, we compute the
gravitational waves emitted by the system. Conventionally, the gravitational radiation is described in terms of the Newman-Penrose scalar $\Psi_{4}$, which can be expanded in terms of spin-weighted $s=-2$ spherical harmonics~\cite{rezbish,brugman}
\begin{equation}
	r \Psi_4 (t,r,\theta,\phi) = \sum_{l,m} C_{l,m}(t,r) \, Y^{-2}_{l,m} (\theta,\phi)~.
	\label{eq:psi4}
\end{equation}
The coefficients $C_{l,m}$ are extracted from spherical surfaces at different extraction radii, $r_\mathrm{ext}=(150,300,450)\, \rm{km}$.


\begin{figure*}[ht]
	\includegraphics[width=18cm]{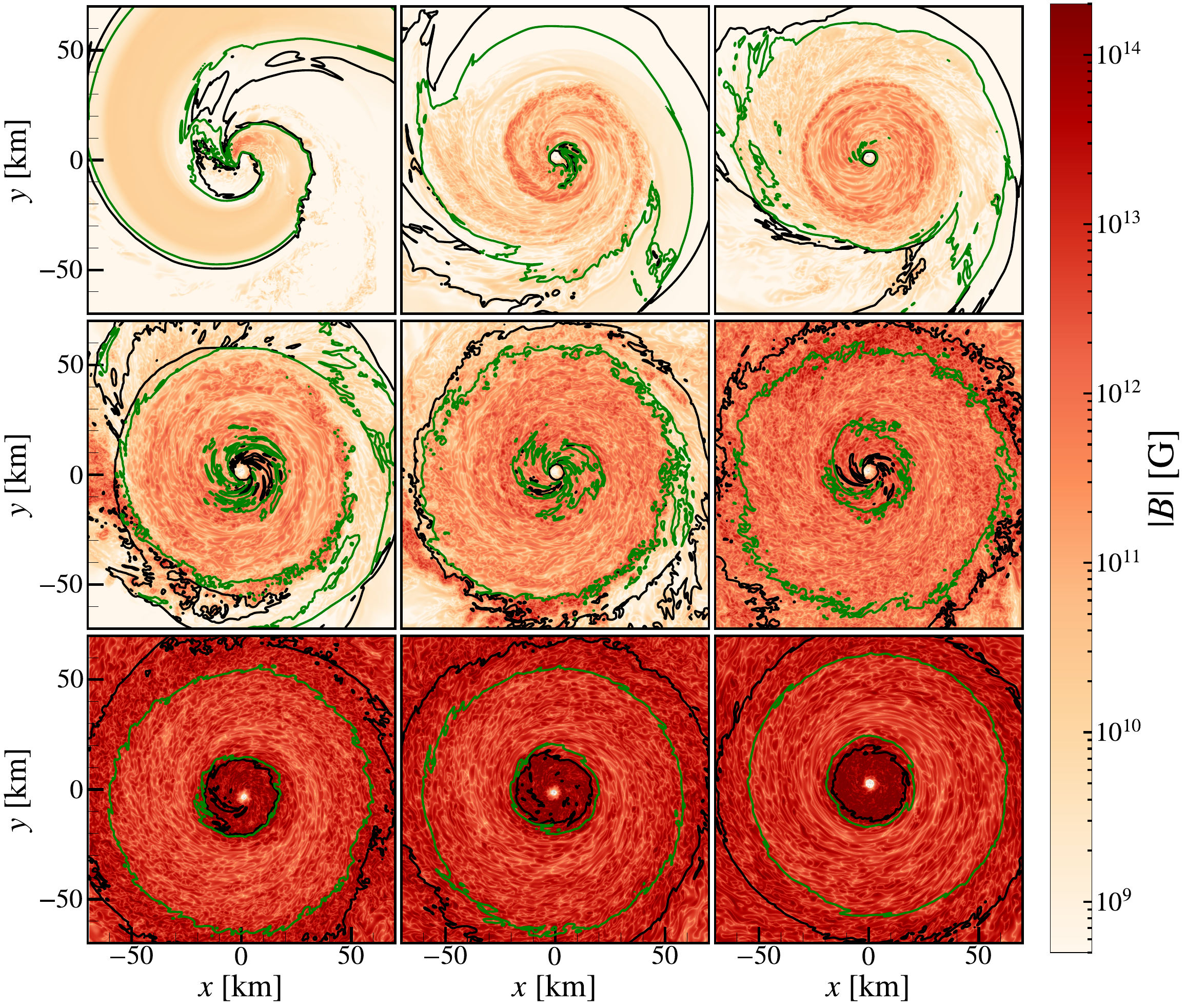}
	\caption{\emph{Evolution of the magnetic field in the orbital plane}. Snapshots of the magnetic field magnitude $|B|$ at times $t = (0.5, 3, 4, 5.5, 7, 9, 16, 20, 25)$ ms with density contour lines corresponding to $\rho = 2\times 10^{10} \text{g cm}^{-3}$ (black) and $\rho = 5\times 10^{10} \text{g cm}^{-3}$ (green). The magnetic field is strongly amplified due to the KHI in the strong shear layers appearing during the accretion disk formation. 
		The KHI is particularly evident in the second and third frames as the spiral regions of magnetic field growth migrate outward in the forming disk.
		\label{BGauss}}
\end{figure*}

\section{Results}
\label{sec:results}
In this section, we present the results of our binary BHNS simulations. First, we describe  qualitatively the dynamics focusing on the evolution of the magnetic field. Then, we study various global quantities, the radial profile of different physical fields in the accretion disk, and the spectra. Finally, we investigate the gravitational waves produced by these mergers. As explained in the previous section, we evolve three different resolutions. Unless otherwise noted, the results shown are those of the medium resolution (i.e., \texttt{MR}). For all three resolutions we have adopted a conservative approach such that the LES terms are applied only above a threshold density $\rho \geq 3\times 10^{10} \text{g cm}^{-3}$, such that they are applied in the densest parts of the remnant.

\subsection{Qualitative dynamics}

The binary performs almost two orbits before the BH tidally disrupts the NS. The qualitative dynamics afterwards can be followed in Fig.~\ref{RHOF}, where time snapshots of the density in the equatorial plane are displayed. Most of the NS is swallowed by the BH on a short time scale of a few milliseconds, except the matter located more distant from the BH, which forms a one-arm spiral structure. The fluid in the outermost region of this one-arm spiral is unbounded and becomes dynamical ejecta. The bound component winds around the BH, eventually interacting with its own tidal tail, forming a turbulent, compact accretion disk with a torus-like shape in $t \lesssim\,10 \,\text{ms}$. 
Subsequently, the disk becomes less turbulent and more axisymmetric, rapidly rotating around the remnant spinning BH.

We are particularly interested in the evolution of the magnetic field, whose evolution is displayed in Fig.~\ref{BGauss}. The Kelvin-Helmholtz instability~(KHI) develops in the contact interfaces of the one-arm spiral, when the head collides and interacts with the tail of the spiral structure.
Shear motion produces vortices in the fluid, which interact non-linearly and induce turbulent dynamics. During this stage, the magnetic field is continuously twisted by the fluid motion, efficiently converting kinetic into magnetic energy. 
As a result, the turbulent dynamo is very active, meaning that it induces a clear exponential growth of the magnetic field, roughly until the formation of the axisymmetric accretion disk at $t \lesssim\,20$ms (see Fig.~\ref{RHOF} and Fig.~\ref{energies_averages}).
The combination of the KHI, the turbulent dynamo, and the winding mechanism (which is likely the least significant
at this early stage) amplify the magnetic field strength $|B|$ from $10^{10}\, \text{G}$ to at least $10^{14} \, \text{G}$ in $t \lesssim 20\, \text{ms}$ after the merger.
Notice that this saturation value is just a lower limit, as higher resolution simulations will accurately resolve higher wave numbers. An upper limit may be roughly estimated by assuming equipartition between kinetic and magnetic energies. From Fig.~\ref{energies_averages}, rescaling the magnetic field  at $t=25$~ms by a factor of roughly $30$ would push the magnetic energy up to the asymptotic kinetic energy, corresponding to a magnetic field of $\approx 3 \times 10^{15}$~G\footnote{{Looking ahead to the spectra in Fig.~\ref{spectra_time}, a similar value for the magnetic field can be found by looking at the peak of the magnetic spectrum.}}.

Once the shear layers have been dissipated and the KHI is no longer active,  the turbulence slowly decays in the disk and other instabilities and amplification mechanisms might affect the magnetic field dynamics. The disk's rotation profile demonstrates differential rotation with a monotonically decreasing angular velocity consistent with the conditions required by the magneto-rotational instability~(MRI).
On timescales longer than those of our simulations, one expects that the MRI could still maintain a certain level of small-scale turbulence.

\begin{figure}[ht]
	\includegraphics[width=8.5cm]{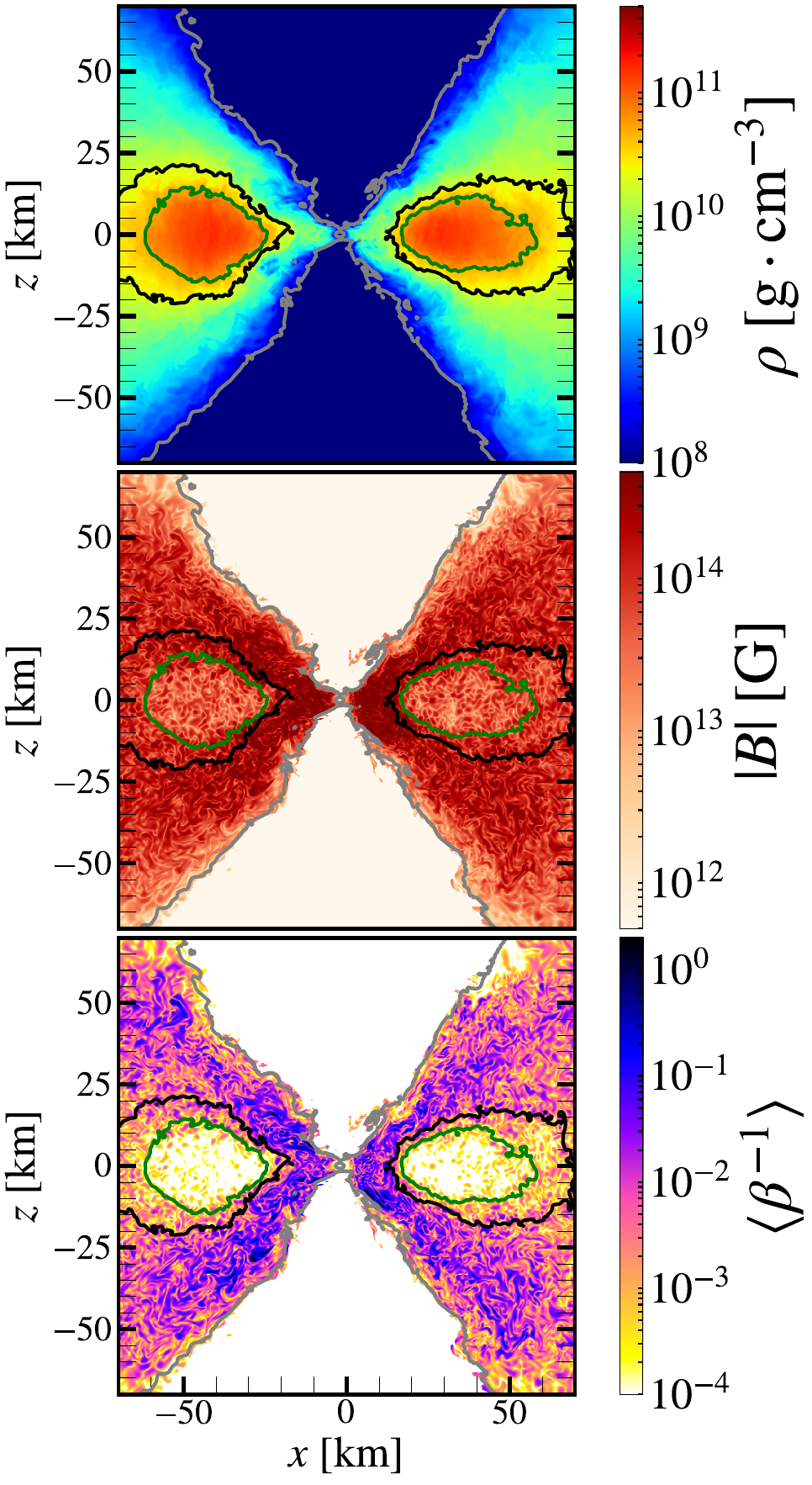}
	\caption{\emph{Late time meridional profile of the disk}.
                 Snapshots of the rest-mass density \textbf{(top)}, the magnetic field strength \textbf{(middle)}, and the inverse of $\beta$ \textbf{(bottom)} at $t = 25\, \text{ms}$, with density contour lines corresponding to $\rho = 10^{8} \text{g cm}^{-3}$ (grey), $\rho = 2\times 10^{10} \text{g cm}^{-3}$ (black), and $\rho = 5\times 10^{10} \text{g cm}^{-3}$ (green). %
        The near bilateral symmetry reflects the disk's approach to axisymmetry.
	The magnetic field is stronger in the innermost region of the envelope, but is sub-dominant in the bulk of the disk. \label{meridional}}
\end{figure}

At late times, the disk tends to an axisymmetric configuration,
which can be seen in the circularized disk in the last couple frames of
Figs.~\ref{RHOF} and~\ref{BGauss} as well as the near symmetry shown in
the meridional profiles of the disk shown in Fig.~\ref{meridional}.
The disk has a torus-like shape with three distinct regions: a high density, compact region forming the \textit{bulk} of the disk (roughly the region $\rho \geq 2\times 10^{10} \text{g cm}^{-3}$ enclosed within the black contour line), an extended corona or \textit{envelope} with low density fluid 
(roughly the region $10^{8} \text{g cm}^{-3} \leq \rho \leq 2 \times 10^{10} \text{g cm}^{-3}$ enclosed between the grey and the black contour lines), and 
then a very rarefied cone-shaped \textit{funnel} with $\rho \leq 10^{8} \text{g cm}^{-3}$
extending roughly $45^o$ around the vertical $z$-axis.
The magnetic field is strongest within the innermost part of the envelope and is much weaker within the bulk.
%

\subsection{Global quantities}

\begin{figure}[h]
	\includegraphics[width=8.5cm]{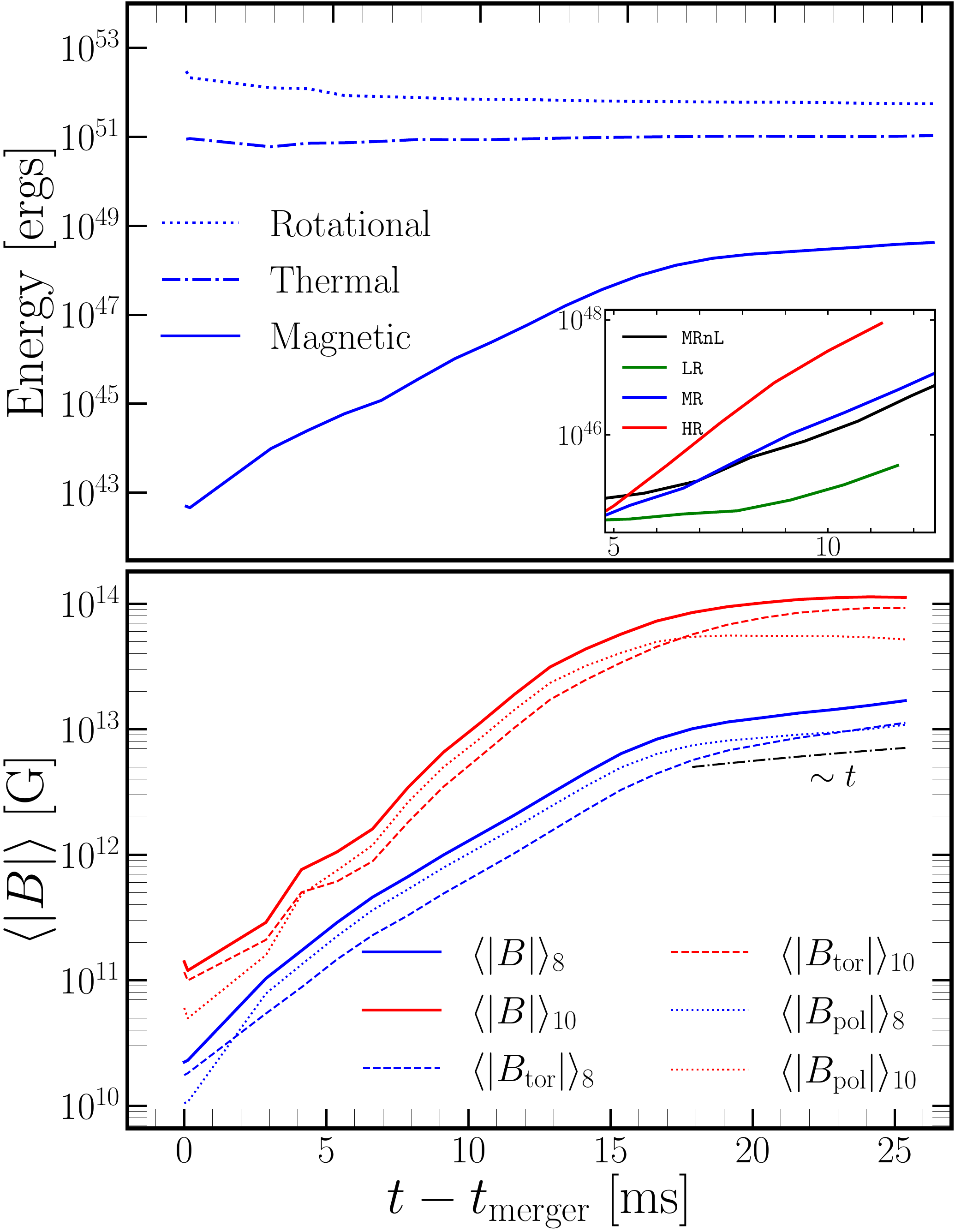}
	\caption{\emph{ Magnetic field evolution.}
          \textbf{Top:} Rotational (dotted), thermal (dashed), and magnetic (solid) energies, integrated over the whole simulation domain. The magnetic energy grows more than six orders of magnitude during the amplification phase. 
As shown in the \textbf{inset}, the growth rate remains strongly dependent on the grid resolution. \textbf{Bottom:} Evolution of the averaged strength of the magnetic field (solid), as well as its toroidal  (dashed) and poloidal (dotted) components, both in the bulk and envelope regions of the disk remnant. The magnetic field is amplified to values $\approx 10^{14}\, \text{G}$ during the amplification phase due to isotropic, turbulent small-scale dynamo, which keeps the poloidal and toroidal components comparable. Notice that the toroidal average presents roughly a linear growth (black dashed-point line) after $t \gtrsim 20\,\text{ms}$.
            \label{energies_averages}}
\end{figure}

A more quantitative study can be obtained by computing relevant analysis quantities and integrating them either over the whole domain or in specific regions. In the top panel of Fig.~\ref{energies_averages} the rotational, thermal, and magnetic energies are shown as functions of time
from the merger. Note that the magnetic energy is initially very small, corresponding to the realistic magnetic fields $|B| \leq 10^{11}\,\text{G}$ for an old neutron star in a binary system. Despite 
beginning with less than $10^{43}$~ergs,
the magnetic energy grows beyond $10^{48}\, \rm{ergs}$ in less than $20\, \rm{ms}$ due to the turbulent small-scale dynamo induced by the MHD instabilities. In the inset of Fig.~\ref{energies_averages}, the strong dependence of the magnetic amplification on the grid resolution is demonstrated, suggesting that significant resolution is still needed to capture accurately these MHD instabilities.
We can also observe in the medium resolution case that the impact of the LES combined with the sub-grid scale gradient model is rather mild.
The reasons for such moderate effects are further discussed in Section~\ref{sec:discussion}.

The magnetic field evolution can be further analyzed by computing its averaged magnitude, together with its toroidal and poloidal components, as displayed in the bottom panel of Fig.~\ref{energies_averages}. The magnetic field strength reaches average values $\approx 10^{14}\, \rm{G}$ in the dense regions with $ \rho \geq 2\times 10^{10} \text{g cm}^{-3}$ (i.e., the disk bulk), but only $\approx 10^{13}\, \text{G}$ when it is averaged in the extended region with densities above $10^{8} \text{g cm}^{-3}$. Interestingly, the poloidal and toroidal averages are comparable and grow at the same rate during the strongest amplification early phase $t \lesssim 20\,\text{ms}$. This equipartition between the two components is another indication that isotropic,
turbulent dynamics are being induced during this stage of the disk formation. After this period, in the envelope of the disk, the toroidal average keeps growing almost linearly with time while the poloidal average remains roughly constant, suggesting that the winding is the dominant amplification mechanism.

\subsection{Radial dependence}

\begin{figure}[t]
	\includegraphics[width=8cm]{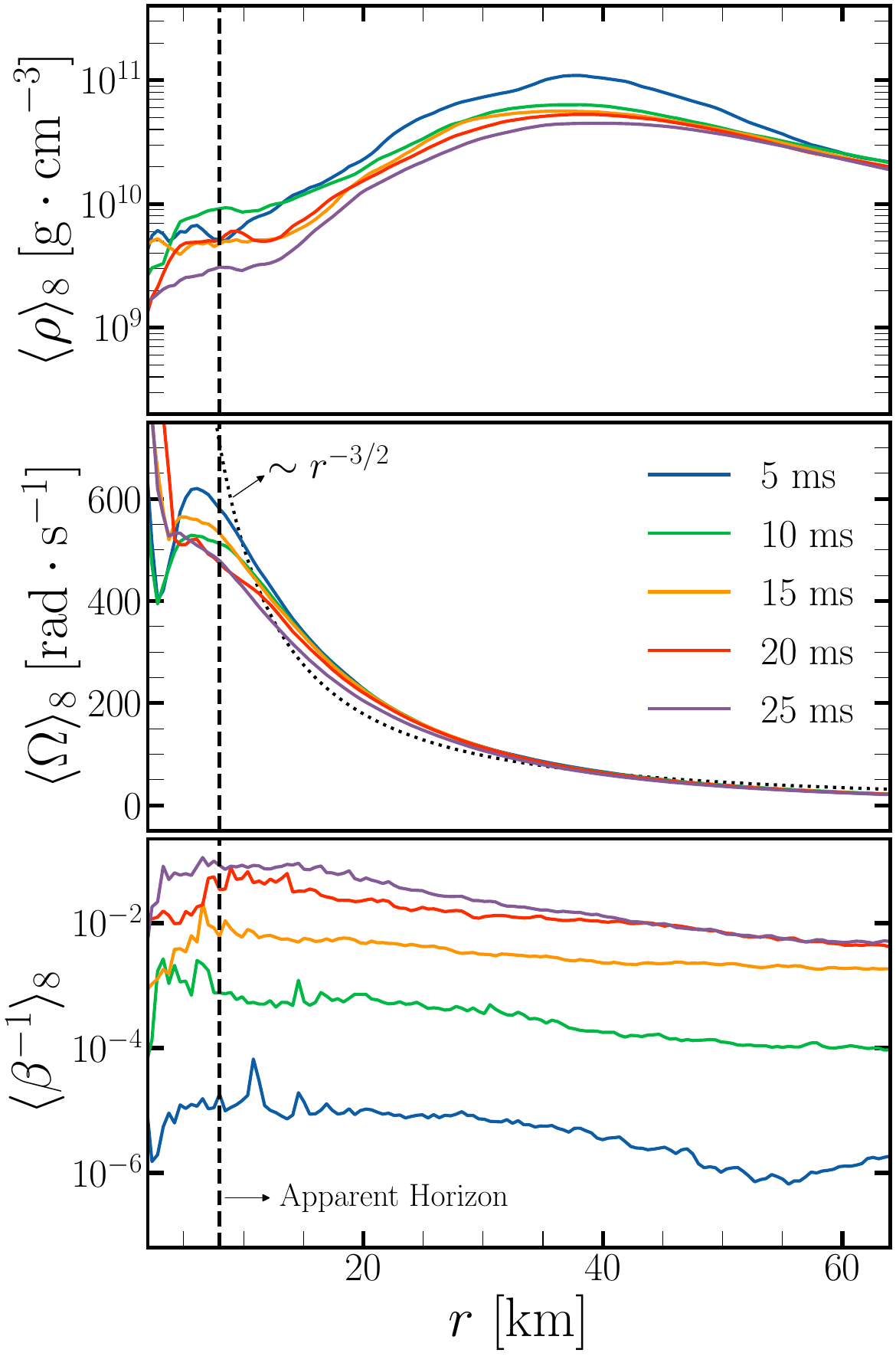}
	\caption{\emph{Cylindrical disk averages.}
The rest-mass density \textbf{(top)}, the angular frequency \textbf{(middle)}, and the ratio $\beta^{-1}$ \textbf{(bottom)} are depicted. A strictly Keplerian angular velocity (dotted) is also shown because the disk angular velocity is nearly Keplerian. The vertical dashed line indicates the approximate location of the BH apparent horizon. {The observed decrease in angular velocity within the disk indicates that the MRI could be active at times longer than the ones evolved here}.
\label{several_cylinders_times}}
\end{figure}

We analyze the radial profile of the accretion disk by computing averages over cylinders centered at the BH, displayed in Fig.~\ref{several_cylinders_times}.
The averaged density, with a maximum around $r \approx 35\, \text{km}$, reaches a smooth profile at $t \approx 10\, \text{ms}$, when the accretion disk becomes roughly axisymmetric. Afterward, the density profile slowly decreases due to the accretion onto the BH and the expansion of the disk both in the radial and $z$ directions. The angular velocity peaks very close to the apparent horizon location (i.e., around $8\, \text{km}$) and then decreases monotonically, rough matching a Keplerian profile at distances $r \gtrsim 12\,\text{km}$. Almost the entire disk is then prone to the MRI, which should eventually be the primary driver of the magnetic field dynamics. The small ratio $\beta^{-1}$ shows that, on average, the fluid pressure dominates at all radii for the duration of the  simulation. Notice, however, that the ratio begins to be significant for $t \gtrsim 25\, \text{ms}$ in the innermost region $r \le 20\, \text{km}$, suggesting that the magnetic field might affect the fluid dynamics at later times.

\begin{figure}[h]
	\includegraphics[width=8.5cm]{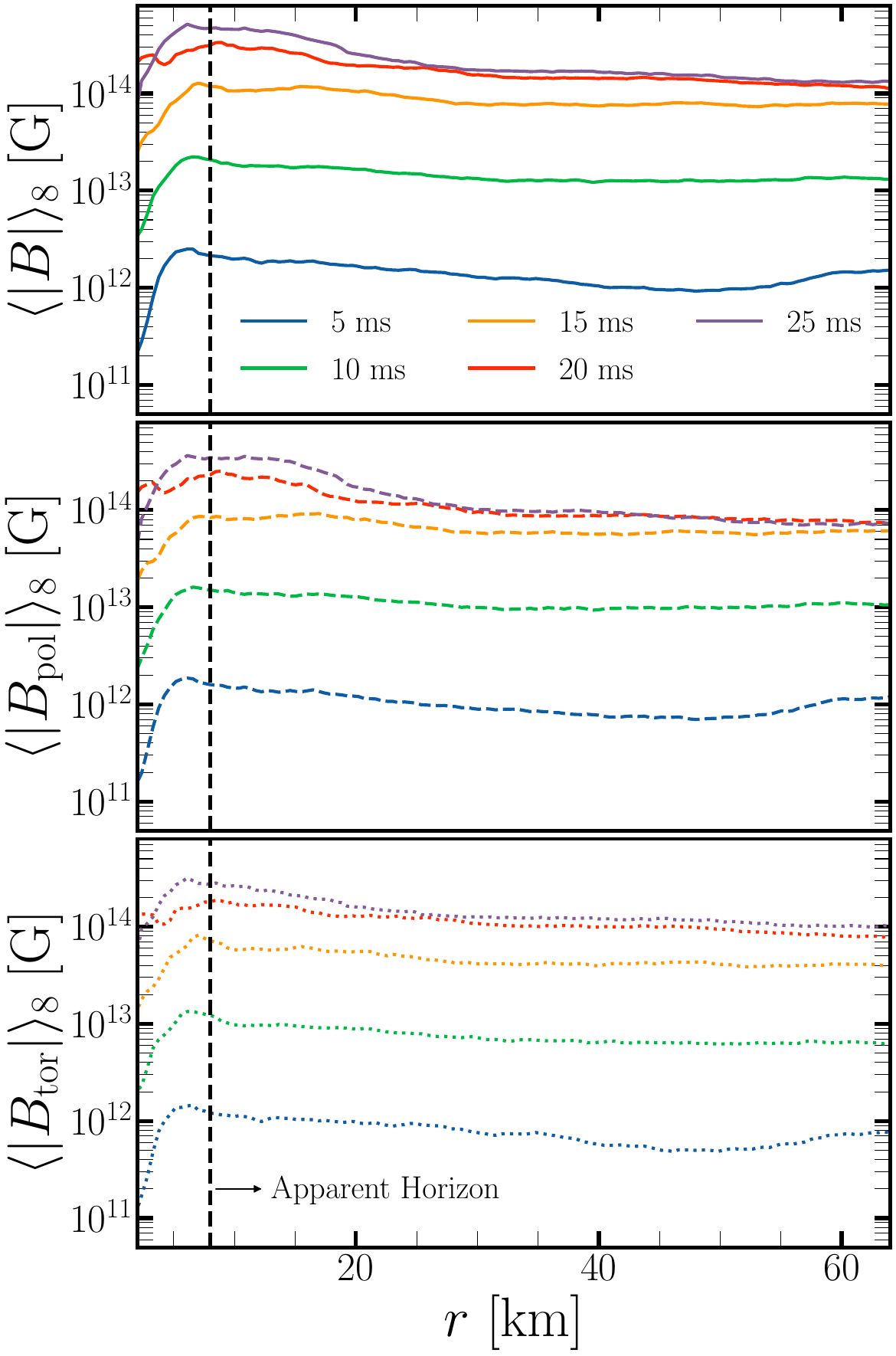}
	\caption{\emph{Evolution of the magnetic field averaged on cylinders}. The average magnetic field strength increases rapidly from $10^{11}\, \text{G}$ to more than $10^{14}\, \text{G}$ in about $20 \, \text{ms}$. The growth is similar for all radii at early times but is mostly suppressed at late time $t \ge 20 \, \text{ms}$ except at the innermost region. \label{avB_cylinders_times}}
\end{figure}

The magnetic field evolution is displayed in Fig.~\ref{avB_cylinders_times}. Certainly, the magnetic field grows rapidly throughout the domain, especially in the innermost part of the disk. At late times $t \gtrsim 20\,\text{ms}$, the growth rate is significantly slowed in that region and is almost completely suppressed in the outermost part. The poloidal and the toroidal components grow simultaneously during the rapid amplification phase, and differences appear only at late times.

\begin{figure}[ht]
	\includegraphics[width=8.5cm]{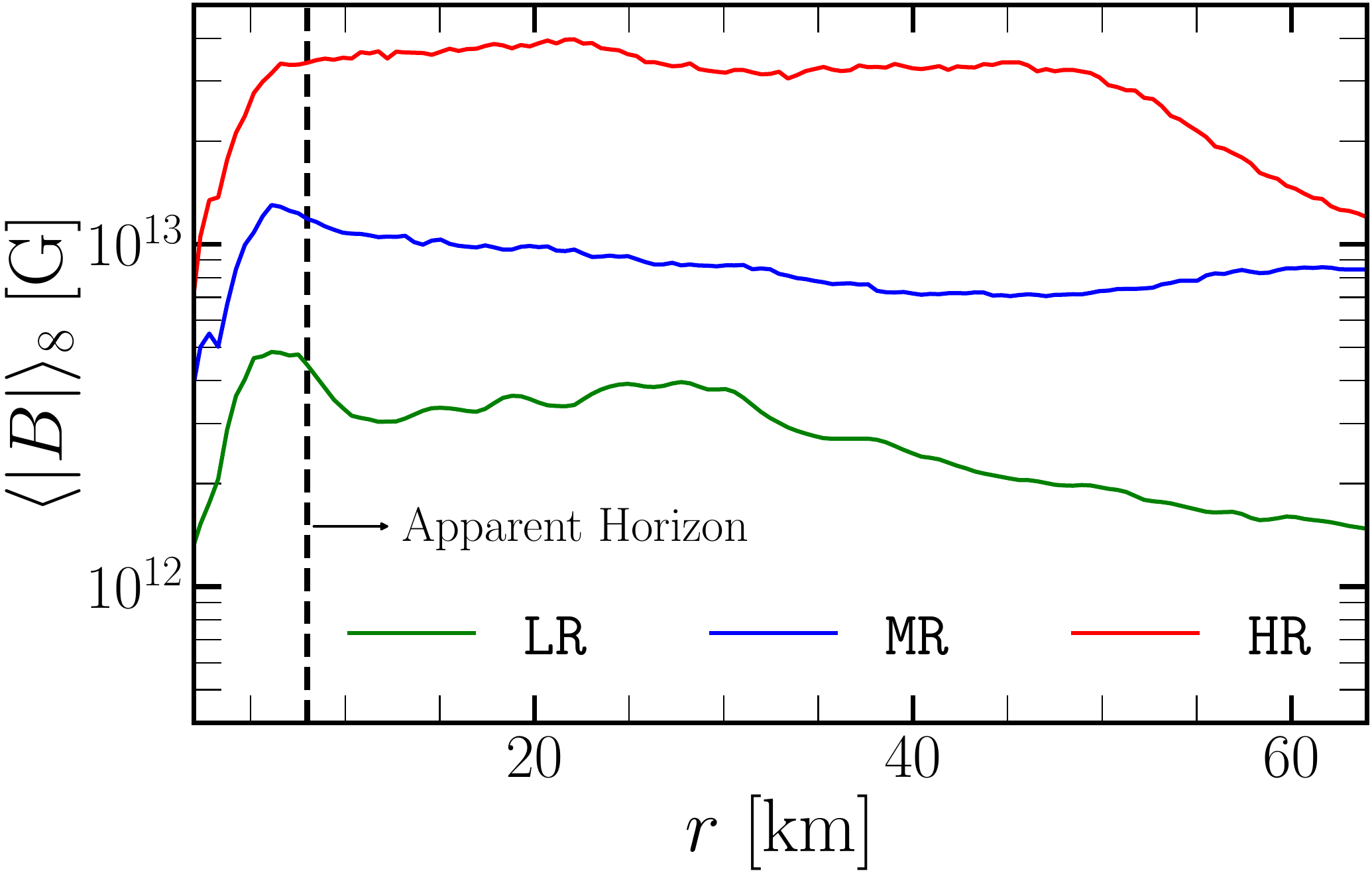}
		
	\caption{\emph{Radial average profile of the magnetic field for different resolutions at \textbf{ $t=9\, \rm{ms}$}.} Both the shape and the strength of the averaged magnetic field differs for the three resolutions, suggesting that more resolution is needed to reach convergence. 
	The poloidal and toroidal profiles behave similarly. %
	\label{avB_cylinders_resolution}}
\end{figure}

Finally, we can examine the effect of spatial resolution on the averaged magnetic field strength, displayed in Fig.~\ref{avB_cylinders_resolution} at {$t=9\, \rm{ms}$} for our three resolutions. Indeed, they differ in shape and amplitude {suggesting that, for these resolutions, the turbulent dynamics is still under-resolved.
}
%

\subsection{Spectra}
It is standard to study the energy spectra in problems displaying turbulent dynamics. Here we follow the same approach.
We take the Fourier transform of the solution within 
a cubic box $[-70\, \text{km} ,70\, \text{km}]^3$ with resolution equal to that of the finest disk resolution of each simulation, $\Delta^\mathrm{disk}_\mathrm{min}$, from Table~\ref{tab:models} (interpolating from coarser grids if needed in order to have a unigrid box).

\begin{figure*}[!ht]
	\includegraphics[width=16cm]{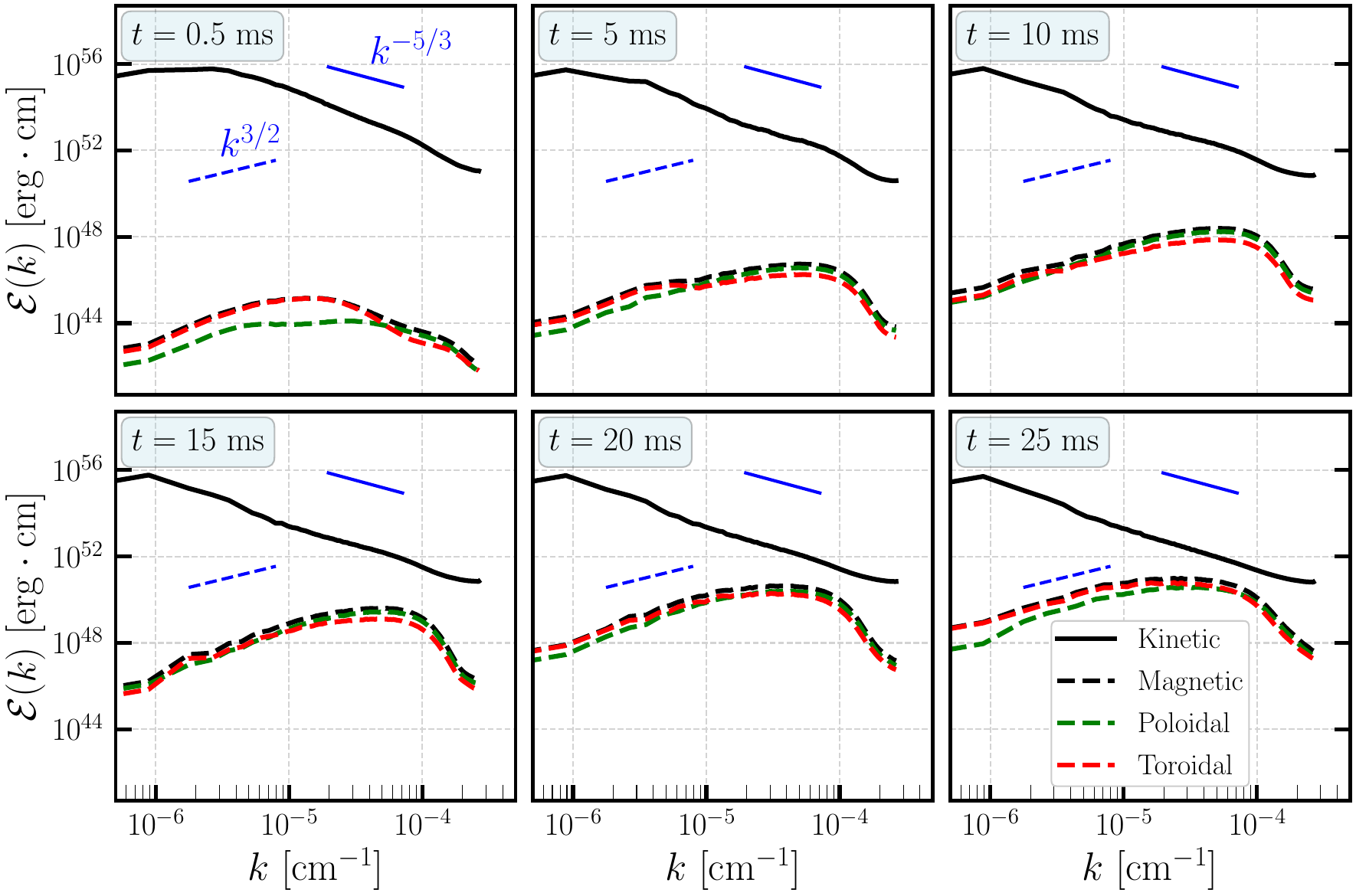}
	\caption{\emph{Evolution of the energy spectra.} Kinetic (solid) and magnetic (dashed) spectra for the medium resolution cases at $t = (0.5, 5, 10, 15, 20, 25)\,\rm{ms}$. The kinetic energy spectra show the standard Kolmogorov power law $k^{-5/3}$ (short solid blue line) in the inertial range. The magnetic energy follows the Kazantsev power law $k^{3/2}$ (short dashed blue line) at low wavenumbers. 
    In the last two snapshots the toroidal magnetic energy grows faster than the poloidal, especially at large scale, suggesting that the winding is responsible for the magnetic amplification at late times.
         \label{spectra_time}}
\end{figure*}

The evolution of the kinetic, Eq.~(\ref{eq:spectra_k}), and magnetic energy, Eq.~(\ref{eq:spectra_b}), spectra are displayed in Fig.~\ref{spectra_time} for our medium resolution case. Interestingly, after a short transient both spectra follow the behavior expected in a turbulent weakly magnetized fluid: the kinetic energy spectrum decreases with the \textit{Kolmogorov power law dependence} $k^{-5/3}$ while the magnetic field spectrum at low wave-numbers follows the \textit{Kazantsev power law} $k^{3/2}$~\cite{1968JETP...26.1031K,1992ApJ...396..606K}. Although the transfer from kinetic to magnetic energy occurs at all scales, it is clearly more effective
at high wavenumbers (i.e., small scales), as expected again in a turbulent regime. Notice also the equipartition between the poloidal and toroidal magnetic spectra, at least up to $t \lesssim 20\, \rm{ms}$, another indicator of an isotropic turbulent regime.

\begin{figure}[t]
	\includegraphics[width=8.5cm]{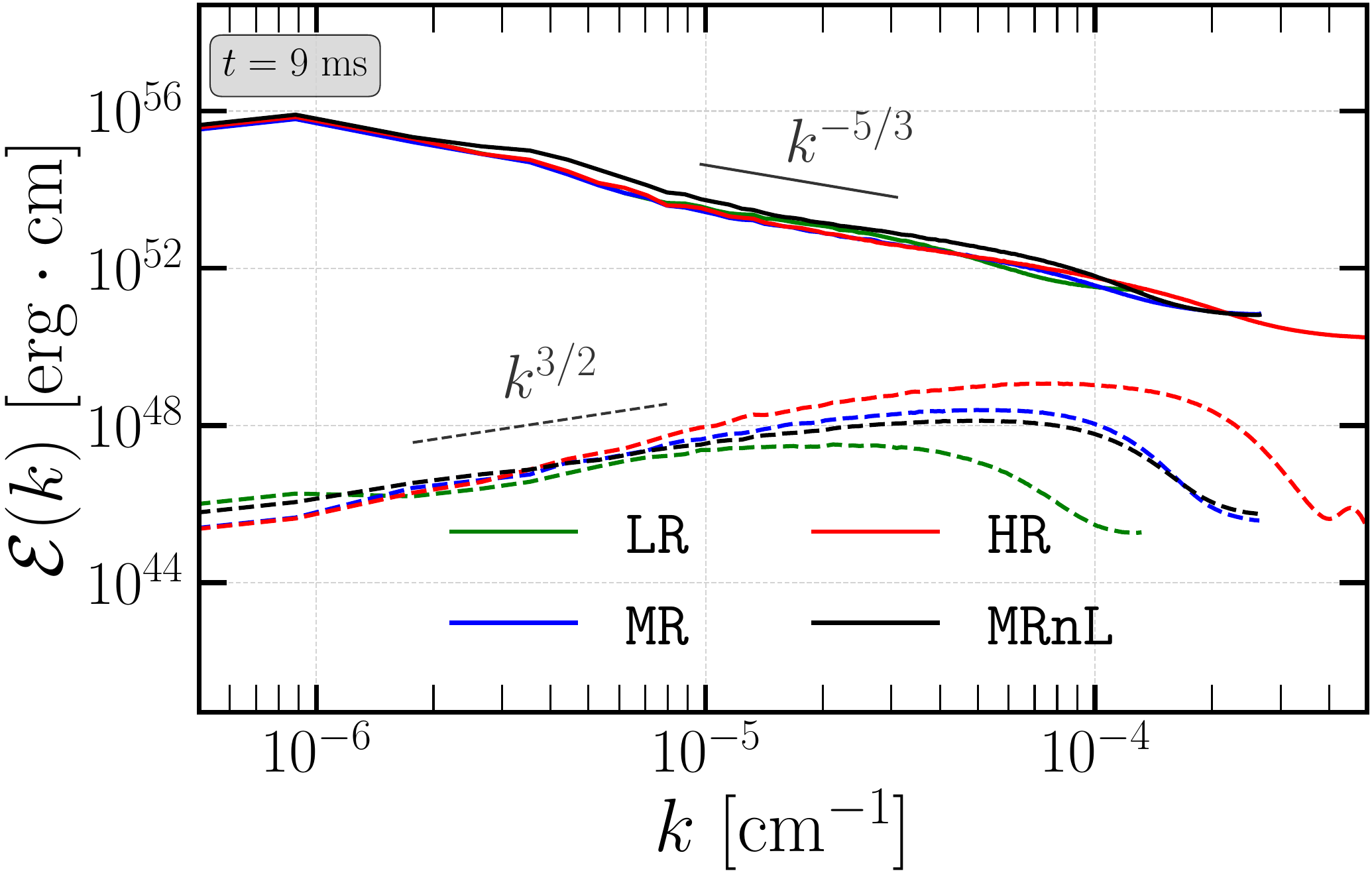}
	\caption{\emph{Energy spectra.} 
        The kinetic~(solid) and magnetic spectra~(dashed) are displayed at $t=9$\,ms for different resolutions. 
	As the resolution increases, the magnetic energy reaches smaller scales (larger wavenumber).
	In contrast, the kinetic energy spectrum is much less sensitive to resolution, suggesting that the fluid motion may be mostly converged at these resolutions. 
        \label{spectra_resolution}}
\end{figure}

We can also study the effects of resolution in the spectra. The spectra for the three different resolutions, along with a case without LES, are displayed in Fig.~\ref{spectra_resolution} at $t=9\, \text{ms}$. The impact of the resolution is quite significant, confirming again that the magnetic energy amplification is more effective at higher resolutions because the MHD instabilities operate more strongly at the smallest resolved scales. 
However, it is important to stress that 
at low and intermediate wave-numbers the slopes of the magnetic energy spectra obtained with our three resolutions is consistent with the expected Kazantsev power law, suggesting that these scales are captured correctly in our simulations. 

\begin{figure}[h]
	\includegraphics[width=8.5cm]{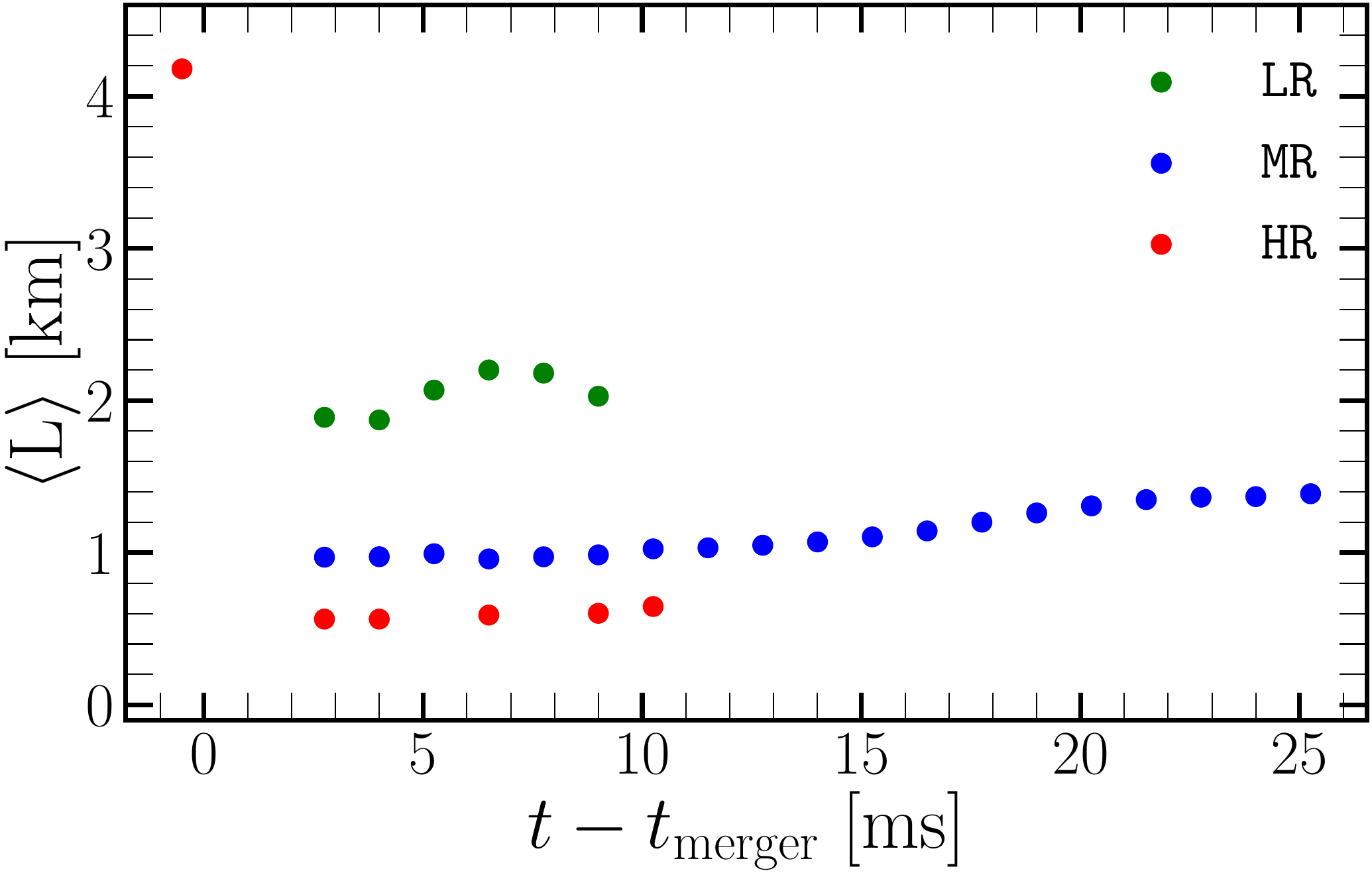}
	\caption{\emph{Evolution of the characteristic length of the magnetic field.} The time evolution of the characteristic length of the magnetic field (see Eq.~\eqref{eq:coherence}). The first point corresponds to the late inspiral, before the disruption. After disk formation, the coherence scale grows monotonically with time, slowing down significantly for $t \gtrsim 20\,\text{ms}$. \label{characteristic_scale}
         }
\end{figure}

We can estimate the coherent, or characteristic, spatial scale of the magnetic field by using the spectra-weighted average wave-number, Eq.~\eqref{eq:coherence}. The evolution of this characteristic scale $\langle L \rangle$ is displayed in Fig.~\ref{characteristic_scale} for our three LES resolutions. 
The difference in average scales decreases as the resolution is increased, (i.e., $\langle L \rangle_{\texttt{LR}} - \langle L \rangle_{\texttt{MR}} \approx 1\,\text{km}$ while $\langle L \rangle_{\texttt{MR}} - \langle L \rangle_{\texttt{HR}} \approx 0.35\, \text{km}$). {Although other magnetic quantities show no clear sign of convergence, that this scale $\langle L \rangle$ tends toward an asymptotic value suggests that we may be close to the convergent regime.}

\subsection{Gravitational waves}

\begin{figure}[t]
	\includegraphics[width=8.5cm]{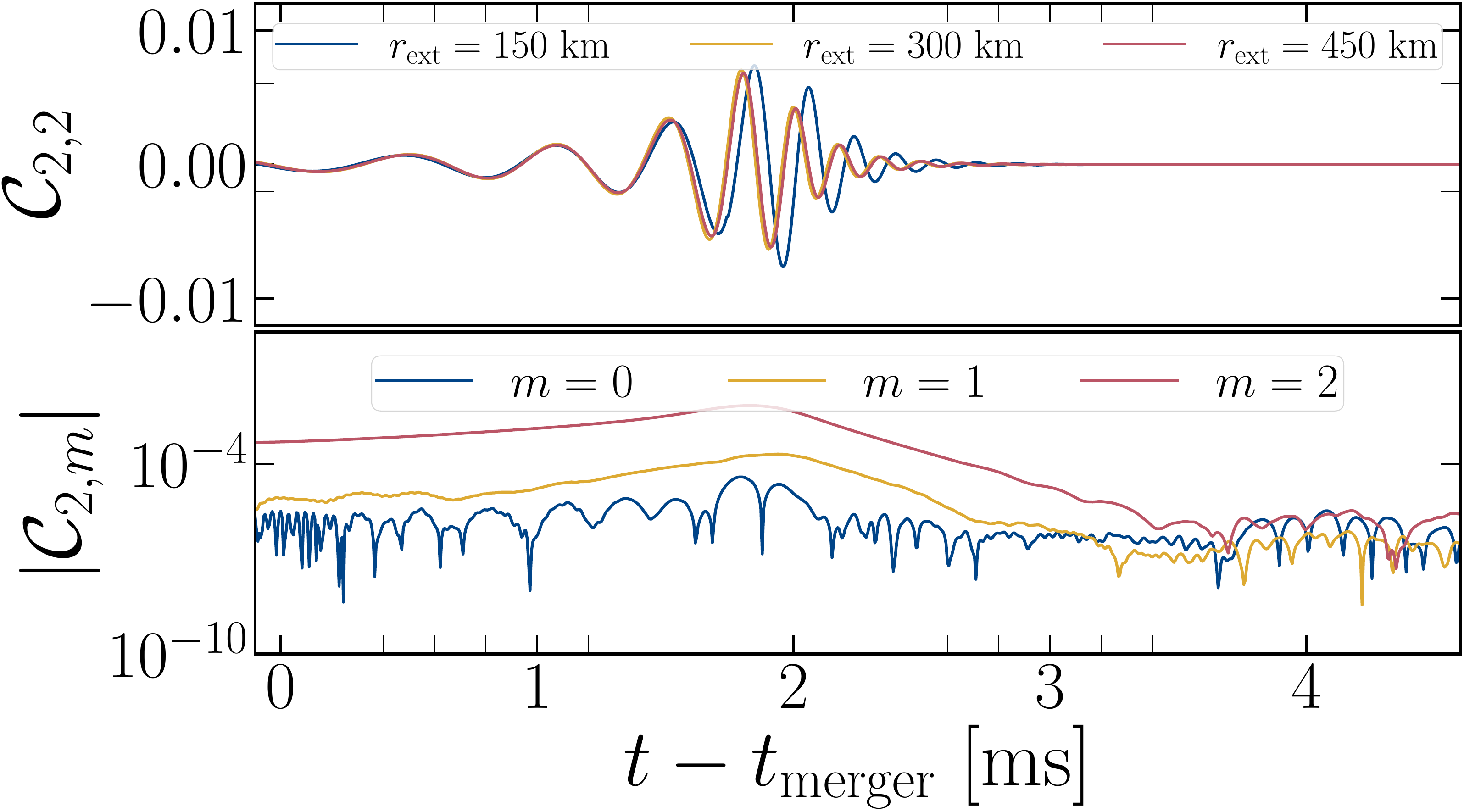}
	\caption{\emph{Gravitational waves}. \textbf{Top:} Dominant $l=m=2$ mode of the Newman-Penrose scalar $\Psi_4$, as a function of time, computed at three spherical extraction surfaces located with radii $r_\mathrm{ext}=(150,300,450)\,\rm{km}$. 
    The signals extracted from the two larger spheres match quite well.
      \textbf{Bottom:} The modulus $|C_{l=2,m}|$ of the $\Psi_4$ modes corresponding to $l=2$. The $m=2$ mode dominates during the full coalescence. \label{GW}} 
\end{figure}

The %
gravitational wave signal for the entire merger is displayed in Fig.~\ref{GW} for the high resolution LES simulation. In the top panel we show the main GW-mode calculated at three different extraction radii in order to assess the error due to finite extraction radius, demonstrating good agreement for $r_\mathrm{ext} \geq 300\,\text{km}$. 
The gravitational wave signal exhibits a structure very similar to the inspiral-merger-ringdown observed in binary black hole mergers due to the dominance of the black hole in all the stages.
As the remnant settles down, the signal relaxes to a simple decaying oscillatory function, which is almost zero within a few milliseconds after merger. 
The merger of a moderately unequal-mass BNS leads naturally to a dominant quadrupole (i.e., $m=2$) and a subdominant dipolar (i.e., $m=1$) gravitational wave mode.
Previous works (see,  e.g.,~\cite{paschalidis2015one,2016PhRvD..94f4011R,east2016relativistic,2016PhRvD..94d3003L}) have studied a non-axisymmetric instability that enhances the $m=1$ mode, which  might even become dominate over the $m=2$ mode at late times. 
Motivated by this, we analyze the different $\left(l=2,m\right)$  modes of the gravitational wave signal, displayed in the bottom panel of Fig.~\ref{GW} for the high resolution simulation. Unlike in the BNS scenario, we observe that no significant enhancements are observed in the $m=1$ mode, probably because the disk's mass is too small (i.e., relative to the final black hole) to produce visible effects in the gravitational waves.
%

\section{Discussion}\label{sec:discussion}

In this paper we have performed simulations using LES with the gradient SGS model to accurately capture the magnetic field dynamics during and after the merger of a black hole with a  neutron star.  These studies are an extension of those performed for binary neutron star systems~\cite{aguilera2020,Aguilera-Miret:2021fre,Palenzuela:2021gdo,2023PhRvD.108j3001A}, where we modified our \mhduet code slightly to handle numerical issues arising with the fluid accreting on the black hole.

We have considered a binary black hole-neutron star system with a mass ratio $q=3$ and a black hole spin $a/M=0.5$, such that the neutron star tidally disrupts before reaching the ISCO and produces a massive torus-like disk accreting onto the final black hole.
Following our previous work with binary neutron stars, we initialized the neutron star's magnetic field with a peak strength of $10^{11}\,\rm{G}$, a realistic upper bound for Gyr-old neutron stars like the ones commonly found in binary systems. This field represents a tiny fraction of the total energy of the system and is much smaller than commonly used in current numerical simulations.  

Our simulations show that, shortly after tidal disruption, the magnetic field is first amplified by the KHI. A shear layer in the one-arm spiral, as it interacts with itself while being wound around the final black hole, is prone to unstable modes that develop vortices at all scales. Turbulent dynamics, driven mainly by this instability, amplifies the magnetic field, which is then redistributed throughout the torus by the fluid flow. A quasi-stationary accretion disk, maintaining some degree of small-scale turbulence, is achieved after approximately $10\,\rm{ms}$. The average magnetic field strength at the saturation phase is approximately $10^{14}\,\rm{G}$. As expected from isotropic turbulence, the poloidal and toroidal components of the magnetic field have comparable strengths at saturation. At later times, the magnetic field grows linearly due to the winding mechanism. 
Eventually, small-scale dynamo induced by the MRI is expected to dominate the magnetic field amplification. 

The behavior just mentioned can be observed clearly in the spectra, which shows also that the magnetic spectrum density tends toward equipartition with the kinetic density at small scales. 
The kinetic energy spectrum follows the expected Kolmogorov power low $k^{-5/3}$ in the inertial range.  On the other hand, the magnetic energy spectra displays the expected Kazantsev power law $k^{3/2}$ for large scales, with a typical coherence length $\langle L \rangle \gtrsim 1\,\rm{km}$ after $10\,\text{ms}$. 

This work is a natural extension of the application of LES techniques to BNS mergers, but, in contrast, the magnetic field amplification does not show clear convergence due to two main differences: the spatial resolution employed in both simulations and, related to this, the mild effect of the LES in BHNS compared with that observed in BNS simulations.
As noted earlier in the text, our finest grid spacing covering the disk is $\Delta^\mathrm{disk}_\mathrm{min} = 120\,\mathrm{m}$. Although among the highest considered for BHNS mergers,
this grid spacing does not reach the grid spacing at which our BNS simulations displayed clearly convergent magnetic field amplification. Obviously, it is easier to perform high resolution simulations of a BNS with a  remnant star of size $\approx \left(20\,\mathrm{km}\right)^3$ than 
of a BHNS in which the accretion disk's size is $\approx \left(100\,\mathrm{km}\right)^3$. 

Related to the lack of convergence is
the mild effect of the LES, most apparent in comparing the \texttt{MRnL} and \texttt{MR} cases which led to similar results. That the resolution is significantly less than that used with BNS mergers likely helps to explain this mild effect. In other words, the LES technique requires sufficient resolution to resolve some of the turbulence spectrum, so that it can effectively model the subgrid scales. In this first work on BHNS, we have adopted a conservative approach such that the LES terms are applied only above a threshold density. Hence, their effects on the resolved dynamics are restricted to the densest parts of the disk, 
and not in the envelope, which might attenuate the global impact of the LES on the resolved dynamics.

Further studies are needed in order to fully understand the amplification process and the impact of magnetic field in the dynamics of the remnant accretion disk. Convergence tests, with even higher resolution and where the SGS terms are active also in the less dense part of the disk, will be needed to discern whether we can capture the small-scale MHD processes and follow their development until, ultimately, the possible formation of a magnetically-dominated jet.

\subsection*{Acknowledgements}

We thank Ricard Aguilera-Miret for useful discussions on the analysis of the results. We are grateful to Samuel Tootle for his kind assistance during the implementation of FUKA. MRI thanks financial support PRE2020-094166 by MCIN/AEI/PID2019-110301GB-I00 and by “FSE invierte en tu futuro”. This work was supported by the Grant PID2022-138963NB-I00 funded by MCIN/AEI/10.13039/501100011033/FEDER, UE. 
This work was also supported by the National Science Foundation via grants PHY-2011383, PHY-2308861.
The authors thankfully acknowledges RES resources provided by BSC in MareNostrum to RES-AECT-2023-2-0002 and RES-AECT-2024-1-0010. This work used the DiRAC@Durham facility managed by the Institute for Computational Cosmology on behalf of the STFC DiRAC HPC Facility (www.dirac.ac.uk). The equipment was funded by BEIS capital funding via STFC capital grants ST/P002293/1, ST/R002371/1 and ST/S002502/1, Durham University and STFC operations grant ST/R000832/1. DiRAC is part of the National e-Infrastructure. This research used Frontera at the Texas Advanced Computing Center, made possible by National Science Foundation award OAC-1818253, and Expanse via
the Advanced Cyberinfrastructure Coordination Ecosystem: Services \& Support (ACCESS) program, which is supported by National Science Foundation grants 2138259, 2138286, 2138307, 2137603, and 2138296.

\clearpage
\bibliographystyle{utphys}
\bibliography{biblio}

\providecommand{\href}[2]{#2}\begingroup\raggedright\begin{thebibliography}{10}

\bibitem{Abbott:2023a}
{\bfseries LIGO Scientific Collaboration, Virgo Collaboration, and KAGRA
  Collaboration} Collaboration, A.~T. D. A. F. e.~a. Abbott, R., ``Population
  of merging compact binaries inferred using gravitational waves through
  gwtc-3,'' \href{http://dx.doi.org/10.1103/PhysRevX.13.011048}{{\em Phys. Rev.
  X} {\bfseries 13} (Mar, 2023) 011048}.
  \url{https://link.aps.org/doi/10.1103/PhysRevX.13.011048}.

\bibitem{Gupta:2023evt}
I.~Gupta, S.~Borhanian, A.~Dhani, D.~Chattopadhyay, R.~Kashyap, V.~A. Villar,
  and B.~S. Sathyaprakash, ``{Neutron star-black hole mergers in next
  generation gravitational-wave observatories},''
  \href{http://dx.doi.org/10.1103/PhysRevD.107.124007}{{\em Phys. Rev. D}
  {\bfseries 107} no.~12, (2023) 124007},
  \href{http://arxiv.org/abs/2301.08763}{{\ttfamily arXiv:2301.08763 [gr-qc]}}.

\bibitem{Broekgaarden:2022}
F.~S. Broekgaarden, E.~Berger, S.~Stevenson, S.~Justham, I.~Mandel,
  M.~Chruślińska, L.~A.~C. van Son, T.~Wagg, A.~Vigna-Gómez, S.~E.
  de Mink, D.~Chattopadhyay, and C.~J. Neijssel, ``{Impact of massive binary
  star and cosmic evolution on gravitational wave observations – II. Double
  compact object rates and properties},''
  \href{http://dx.doi.org/10.1093/mnras/stac1677}{{\em Monthly Notices of the
  Royal Astronomical Society} {\bfseries 516} no.~4, (07, 2022) 5737--5761}.
  \url{https://doi.org/10.1093/mnras/stac1677}.

\bibitem{Gompertz:2022}
B.~P. Gompertz, M.~Nicholl, P.~Schmidt, G.~Pratten, and A.~Vecchio,
  ``{Constraints on compact binary merger evolution from spin-orbit
  misalignment in gravitational-wave observations},''
  \href{http://dx.doi.org/10.1093/mnras/stac029}{{\em Monthly Notices of the
  Royal Astronomical Society} {\bfseries 511} no.~1, (01, 2022) 1454--1461}.
  \url{https://doi.org/10.1093/mnras/stac029}.

\bibitem{Michaely:2022}
E.~Michaely and S.~Naoz, ``Ultrawide black hole—neutron star binaries as a
  possible source for gravitational waves and short gamma-ray bursts,''
  \href{http://dx.doi.org/10.3847/1538-4357/ac8a92}{{\em The Astrophysical
  Journal} {\bfseries 936} no.~2, (Sep, 2022) 184}.
  \url{https://dx.doi.org/10.3847/1538-4357/ac8a92}.

\bibitem{Foucart:2020ats}
F.~Foucart, ``{A brief overview of black hole-neutron star mergers},''
  \href{http://dx.doi.org/10.3389/fspas.2020.00046}{{\em Front. Astron. Space
  Sci.} {\bfseries 7} (2020) 46},
  \href{http://arxiv.org/abs/2006.10570}{{\ttfamily arXiv:2006.10570
  [astro-ph.HE]}}.

\bibitem{Kyutoku:2021icp}
K.~Kyutoku, M.~Shibata, and K.~Taniguchi, ``{Coalescence of black
  hole\textendash{}neutron star binaries},''
  \href{http://dx.doi.org/10.1007/s41114-021-00033-4}{{\em Living Rev. Rel.}
  {\bfseries 24} no.~1, (2021) 5},
  \href{http://arxiv.org/abs/2110.06218}{{\ttfamily arXiv:2110.06218
  [astro-ph.HE]}}.

\bibitem{mhduet_webpage}
``\texttt{MHDuet} website: A distributed {AMR}, {GRMHD} code with {LES} and
  neutrinos,'' 2024.
\newblock \url{http://mhduet.liu.edu/}.

\bibitem{Zhiyin:2015}
Y.~Zhiyin, ``Large-eddy simulation: Past, present and the future,''
  \href{http://dx.doi.org/https://doi.org/10.1016/j.cja.2014.12.007}{{\em
  Chinese Journal of Aeronautics} {\bfseries 28} no.~1, (2015) 11--24}.
  \url{https://www.sciencedirect.com/science/article/pii/S1000936114002064}.

\bibitem{leonard75}
A.~{Leonard}, ``{Energy Cascade in Large-Eddy Simulations of Turbulent Fluid
  Flows},'' \href{http://dx.doi.org/10.1016/S0065-2687(08)60464-1}{{\em
  Advances in Geophysics} {\bfseries 18} (1975) 237--248}.

\bibitem{muller02a}
W.-C. {M{\"u}ller} and D.~{Carati}, ``{Dynamic gradient-diffusion subgrid
  models for incompressible magnetohydrodynamic turbulence},''
  \href{http://dx.doi.org/10.1063/1.1448498}{{\em Physics of Plasmas}
  {\bfseries 9} (Mar., 2002) 824--834}.

\bibitem{grete16}
P.~{Grete}, D.~G. {Vlaykov}, W.~{Schmidt}, and D.~R.~G. {Schleicher}, ``{A
  nonlinear structural subgrid-scale closure for compressible MHD. II. A priori
  comparison on turbulence simulation data},''
  \href{http://dx.doi.org/10.1063/1.4954304}{{\em Physics of Plasmas}
  {\bfseries 23} no.~6, (June, 2016) 062317},
  \href{http://arxiv.org/abs/1606.01573}{{\ttfamily arXiv:1606.01573
  [physics.flu-dyn]}}.

\bibitem{grete17phd}
P.~{Grete}, {\em {LES of compressible magnetohydrodynamic turbulence}}.
\newblock PhD thesis, Max-Planck-Institut f{\"u}r Sonnensystemforschung, 2017.

\bibitem{vigano19b}
D.~{Vigan{\`o}}, R.~{Aguilera-Miret}, and C.~{Palenzuela}, ``{Extension of the
  subgrid-scale gradient model for compressible magnetohydrodynamics turbulent
  instabilities},'' \href{http://dx.doi.org/10.1063/1.5121546}{{\em Physics of
  Fluids} {\bfseries 31} no.~10, (Oct, 2019) 105102},
  \href{http://arxiv.org/abs/1904.04099}{{\ttfamily arXiv:1904.04099
  [physics.flu-dyn]}}.

\bibitem{carrasco19}
F.~Carrasco, D.~Vigan{\`o}, and C.~Palenzuela, ``Gradient subgrid-scale model
  for relativistic mhd large-eddy simulations,'' {\em Physical Review D}
  {\bfseries 101} no.~6, (2020) 063003.

\bibitem{vigano20}
D.~{Vigan{\`o}}, R.~{Aguilera-Miret}, F.~{Carrasco}, B.~{Mi{\~n}ano}, and
  C.~{Palenzuela}, ``{General relativistic MHD large eddy simulations with
  gradient subgrid-scale model},''
  \href{http://dx.doi.org/10.1103/PhysRevD.101.123019}{{\em \prd} {\bfseries
  101} no.~12, (June, 2020) 123019},
  \href{http://arxiv.org/abs/2004.00870}{{\ttfamily arXiv:2004.00870 [gr-qc]}}.

\bibitem{aguilera2020}
R.~Aguilera-Miret, D.~Vigan{\`o}, F.~Carrasco, B.~Mi{\~n}ano, and
  C.~Palenzuela, ``Turbulent magnetic-field amplification in the first 10
  milliseconds after a binary neutron star merger: Comparing high-resolution
  and large-eddy simulations,'' {\em Physical Review D} {\bfseries 102} no.~10,
  (2020) 103006.

\bibitem{Palenzuela:2021gdo}
C.~Palenzuela, R.~Aguilera-Miret, F.~Carrasco, R.~Ciolfi, J.~V. Kalinani,
  W.~Kastaun, B.~Mi\~nano, and D.~Vigan\`o, ``{Turbulent magnetic field
  amplification in binary neutron star mergers},''
  \href{http://dx.doi.org/10.1103/PhysRevD.106.023013}{{\em Phys. Rev. D}
  {\bfseries 106} no.~2, (2022) 023013},
  \href{http://arxiv.org/abs/2112.08413}{{\ttfamily arXiv:2112.08413 [gr-qc]}}.

\bibitem{Aguilera-Miret:2021fre}
R.~Aguilera-Miret, D.~Vigan\`o, and C.~Palenzuela, ``{Universality of the
  Turbulent Magnetic Field in Hypermassive Neutron Stars Produced by Binary
  Mergers},'' \href{http://dx.doi.org/10.3847/2041-8213/ac50a7}{{\em Astrophys.
  J. Lett.} {\bfseries 926} no.~2, (2022) L31},
  \href{http://arxiv.org/abs/2112.08406}{{\ttfamily arXiv:2112.08406 [gr-qc]}}.

\bibitem{Palenzuela:2022kqk}
C.~Palenzuela, S.~Liebling, and B.~Mi\~nano, ``{Large eddy simulations of
  magnetized mergers of neutron stars with neutrinos},''
  \href{http://dx.doi.org/10.1103/PhysRevD.105.103020}{{\em Phys. Rev. D}
  {\bfseries 105} no.~10, (2022) 103020},
  \href{http://arxiv.org/abs/2204.02721}{{\ttfamily arXiv:2204.02721 [gr-qc]}}.

\bibitem{2023PhRvD.108j3001A}
R.~{Aguilera-Miret}, C.~{Palenzuela}, F.~{Carrasco}, and D.~{Vigan{\`o}},
  ``{Role of turbulence and winding in the development of large-scale, strong
  magnetic fields in long-lived remnants of binary neutron star mergers},''
  \href{http://dx.doi.org/10.1103/PhysRevD.108.103001}{{\em \prd} {\bfseries
  108} no.~10, (Nov., 2023) 103001},
  \href{http://arxiv.org/abs/2307.04837}{{\ttfamily arXiv:2307.04837
  [astro-ph.HE]}}.

\bibitem{Radice:2024gic}
D.~Radice and I.~Hawke, ``{Turbulence modelling in neutron star merger
  simulations},'' \href{http://dx.doi.org/10.1007/s41115-023-00019-9}{{\em Liv.
  Rev. Comput. Astrophys.} {\bfseries 10} no.~1, (2024) 1},
  \href{http://arxiv.org/abs/2402.03224}{{\ttfamily arXiv:2402.03224
  [astro-ph.HE]}}.

\bibitem{Chawla:2010sw}
S.~Chawla, M.~Anderson, M.~Besselman, L.~Lehner, S.~L. Liebling, P.~M. Motl,
  and D.~Neilsen, ``{Mergers of Magnetized Neutron Stars with Spinning Black
  Holes: Disruption, Accretion and Fallback},''
  \href{http://dx.doi.org/10.1103/PhysRevLett.105.111101}{{\em Phys. Rev.
  Lett.} {\bfseries 105} (2010) 111101},
  \href{http://arxiv.org/abs/1006.2839}{{\ttfamily arXiv:1006.2839 [gr-qc]}}.

\bibitem{Etienne:2011ea}
Z.~B. Etienne, Y.~T. Liu, V.~Paschalidis, and S.~L. Shapiro, ``{General
  relativistic simulations of black hole-neutron star mergers: Effects of
  magnetic fields},'' \href{http://dx.doi.org/10.1103/PhysRevD.85.064029}{{\em
  Phys. Rev. D} {\bfseries 85} (2012) 064029},
  \href{http://arxiv.org/abs/1112.0568}{{\ttfamily arXiv:1112.0568
  [astro-ph.HE]}}.

\bibitem{Etienne:2015cea}
Z.~B. Etienne, V.~Paschalidis, R.~Haas, P.~M\"osta, and S.~L. Shapiro,
  ``{IllinoisGRMHD: An Open-Source, User-Friendly GRMHD Code for Dynamical
  Spacetimes},'' \href{http://dx.doi.org/10.1088/0264-9381/32/17/175009}{{\em
  Class. Quant. Grav.} {\bfseries 32} (2015) 175009},
  \href{http://arxiv.org/abs/1501.07276}{{\ttfamily arXiv:1501.07276
  [astro-ph.HE]}}.

\bibitem{Kiuchi:2015qua}
K.~Kiuchi, Y.~Sekiguchi, K.~Kyutoku, M.~Shibata, K.~Taniguchi, and T.~Wada,
  ``{High resolution magnetohydrodynamic simulation of black hole-neutron star
  merger: Mass ejection and short gamma ray bursts},''
  \href{http://dx.doi.org/10.1103/PhysRevD.92.064034}{{\em Phys. Rev. D}
  {\bfseries 92} no.~6, (2015) 064034},
  \href{http://arxiv.org/abs/1506.06811}{{\ttfamily arXiv:1506.06811
  [astro-ph.HE]}}.

\bibitem{Paschalidis:2014qra}
V.~Paschalidis, M.~Ruiz, and S.~L. Shapiro, ``{Relativistic Simulations of
  Black Hole\textendash{}neutron Star Coalescence: the jet Emerges},''
  \href{http://dx.doi.org/10.1088/2041-8205/806/1/L14}{{\em Astrophys. J.
  Lett.} {\bfseries 806} no.~1, (2015) L14},
  \href{http://arxiv.org/abs/1410.7392}{{\ttfamily arXiv:1410.7392
  [astro-ph.HE]}}.

\bibitem{Ruiz:2017due}
M.~Ruiz, S.~L. Shapiro, and A.~Tsokaros, ``{GW170817, General Relativistic
  Magnetohydrodynamic Simulations, and the Neutron Star Maximum Mass},''
  \href{http://dx.doi.org/10.1103/PhysRevD.97.021501}{{\em Phys. Rev. D}
  {\bfseries 97} no.~2, (2018) 021501},
  \href{http://arxiv.org/abs/1711.00473}{{\ttfamily arXiv:1711.00473
  [astro-ph.HE]}}.

\bibitem{10.1093/mnras/stab1824}
E.~R. Most, L.~J. Papenfort, S.~D. Tootle, and L.~Rezzolla, ``{On accretion
  discs formed in MHD simulations of black hole–neutron star mergers with
  accurate microphysics},''
  \href{http://dx.doi.org/10.1093/mnras/stab1824}{{\em Monthly Notices of the
  Royal Astronomical Society} {\bfseries 506} no.~3, (07, 2021) 3511--3526}.
  \url{https://doi.org/10.1093/mnras/stab1824}.

\bibitem{Tauris:2017}
T.~Tauris, M.~Kramer, P.~Freire, N.~Wex, H.~Janka, N.~Langer, P.~Podsiadlowski,
  E.~Bozzo, S.~Chaty, M.~Kruckow, E.~{Van Den Heuvel}, J.~Antoniadis,
  R.~Breton, and D.~Champion, ``Formation of double neutron star systems,''
  \href{http://dx.doi.org/10.3847/1538-4357/aa7e89}{{\em Astrophysical Journal}
  {\bfseries 846} no.~2, (Sept., 2017) }.

\bibitem{Papenfort:2021hod}
L.~J. Papenfort, S.~D. Tootle, P.~Grandcl\'ement, E.~R. Most, and L.~Rezzolla,
  ``{New public code for initial data of unequal-mass, spinning compact-object
  binaries},'' \href{http://dx.doi.org/10.1103/PhysRevD.104.024057}{{\em Phys.
  Rev. D} {\bfseries 104} no.~2, (2021) 024057},
  \href{http://arxiv.org/abs/2103.09911}{{\ttfamily arXiv:2103.09911 [gr-qc]}}.

\bibitem{GRANDCLEMENT20103334}
P.~Grandclément, ``Kadath: A spectral solver for theoretical physics,''
  \href{http://dx.doi.org/https://doi.org/10.1016/j.jcp.2010.01.005}{{\em
  Journal of Computational Physics} {\bfseries 229} no.~9, (2010) 3334--3357}.
  \url{https://www.sciencedirect.com/science/article/pii/S0021999110000203}.

\bibitem{bezares17}
M.~Bezares, C.~Palenzuela, and C.~Bona, ``Final fate of compact boson star
  mergers,'' \href{http://dx.doi.org/10.1103/PhysRevD.95.124005}{{\em Phys.
  Rev. D} {\bfseries 95} (Jun, 2017) 124005}.
  \url{https://link.aps.org/doi/10.1103/PhysRevD.95.124005}.

\bibitem{simf3}
C.~Palenzuela, B.~Miñano, D.~Viganò, A.~Arbona, C.~Bona-Casas, A.~Rigo,
  M.~Bezares, C.~Bona, and J.~Massó, ``A simflowny-based finite-difference
  code for high-performance computing in numerical relativity,'' {\em Classical
  and Quantum Gravity} {\bfseries 35} no.~18, (2018) 185007.
  \url{http://stacks.iop.org/0264-9381/35/i=18/a=185007}.

\bibitem{liebling20}
S.~L. {Liebling}, C.~{Palenzuela}, and L.~{Lehner}, ``Toward fidelity and
  scalability in non-vacuum mergers,''
  \href{http://dx.doi.org/10.1088/1361-6382/ab8fcd}{{\em Classical and Quantum
  Gravity} {\bfseries 37} no.~13, (Jun, 2020) 135006}.
  \url{https://doi.org/10.1088%2F1361-6382%2Fab8fcd}.

\bibitem{alic}
D.~Alic, C.~Bona-Casas, C.~Bona, L.~Rezzolla, and C.~Palenzuela, ``Conformal
  and covariant formulation of the z4 system with constraint-violation
  damping,'' \href{http://dx.doi.org/10.1103/PhysRevD.85.064040}{{\em Phys.
  Rev. D} {\bfseries 85} (Mar, 2012) 064040}.
  \url{https://link.aps.org/doi/10.1103/PhysRevD.85.064040}.

\bibitem{bezpalen}
M.~Bezares, C.~Palenzuela, and C.~Bona, ``Final fate of compact boson star
  mergers,'' \href{http://dx.doi.org/10.1103/PhysRevD.95.124005}{{\em Phys.
  Rev. D} {\bfseries 95} (Jun, 2017) 124005}.
  \url{https://link.aps.org/doi/10.1103/PhysRevD.95.124005}.

\bibitem{Z41}
C.~Bona, T.~Ledvinka, C.~Palenzuela, and M.~\ifmmode \check{Z}\else
  \v{Z}\fi{}\'a\ifmmode~\check{c}\else \v{c}\fi{}ek, ``General-covariant
  evolution formalism for numerical relativity,''
  \href{http://dx.doi.org/10.1103/PhysRevD.67.104005}{{\em Phys. Rev. D}
  {\bfseries 67} (May, 2003) 104005}.
  \url{https://link.aps.org/doi/10.1103/PhysRevD.67.104005}.

\bibitem{Bona:2003qn}
C.~Bona, T.~Ledvinka, C.~Palenzuela, and M.~Zacek, ``{A Symmetry breaking
  mechanism for the Z4 general covariant evolution system},''
  \href{http://dx.doi.org/10.1103/PhysRevD.69.064036}{{\em Phys. Rev. D}
  {\bfseries 69} (2004) 064036},
  \href{http://arxiv.org/abs/gr-qc/0307067}{{\ttfamily arXiv:gr-qc/0307067}}.

\bibitem{Z44}
C.~Bona, C.~Bona-Casas, and C.~Palenzuela, ``Action principle for
  numerical-relativity evolution systems,''
  \href{http://dx.doi.org/10.1103/PhysRevD.82.124010}{{\em Phys. Rev. D}
  {\bfseries 82} (Dec, 2010) 124010}.
  \url{https://link.aps.org/doi/10.1103/PhysRevD.82.124010}.

\bibitem{gundlach}
C.~Gundlach, G.~Calabrese, I.~Hinder, and J.~M. Martín-García, ``Constraint
  damping in the z4 formulation and harmonic gauge,'' {\em Classical and
  Quantum Gravity} {\bfseries 22} no.~17, (2005) 3767.
  \url{http://stacks.iop.org/0264-9381/22/i=17/a=025}.

\bibitem{PhysRevD.104.084010}
S.~V. Chaurasia, T.~Dietrich, and S.~Rosswog, ``Black hole-neutron star
  simulations with the $\mathsf{BAM}$ code: First tests and simulations,''
  \href{http://dx.doi.org/10.1103/PhysRevD.104.084010}{{\em Phys. Rev. D}
  {\bfseries 104} (Oct, 2021) 084010}.
  \url{https://link.aps.org/doi/10.1103/PhysRevD.104.084010}.

\bibitem{Alic:2013xsa}
D.~Alic, W.~Kastaun, and L.~Rezzolla, ``{Constraint damping of the conformal
  and covariant formulation of the Z4 system in simulations of binary neutron
  stars},'' \href{http://dx.doi.org/10.1103/PhysRevD.88.064049}{{\em Phys. Rev.
  D} {\bfseries 88} no.~6, (2013) 064049},
  \href{http://arxiv.org/abs/1307.7391}{{\ttfamily arXiv:1307.7391 [gr-qc]}}.

\bibitem{BM}
C.~{Bona}, J.~{Mass{\'o}}, E.~{Seidel}, and J.~{Stela}, ``{New Formalism for
  Numerical Relativity},''
  \href{http://dx.doi.org/10.1103/PhysRevLett.75.600}{{\em Physical Review
  Letters} {\bfseries 75} (July, 1995) 600--603},
  \href{http://arxiv.org/abs/gr-qc/9412071}{{\ttfamily gr-qc/9412071}}.

\bibitem{alcub}
M.~Alcubierre, B.~Br\"ugmann, P.~Diener, M.~Koppitz, D.~Pollney, E.~Seidel, and
  R.~Takahashi, ``Gauge conditions for long-term numerical black hole
  evolutions without excision,''
  \href{http://dx.doi.org/10.1103/PhysRevD.67.084023}{{\em Phys. Rev. D}
  {\bfseries 67} (Apr, 2003) 084023}.
  \url{https://link.aps.org/doi/10.1103/PhysRevD.67.084023}.

\bibitem{2022PhRvD.106b3008H}
K.~{Hayashi}, S.~{Fujibayashi}, K.~{Kiuchi}, K.~{Kyutoku}, Y.~{Sekiguchi}, and
  M.~{Shibata}, ``{General-relativistic neutrino-radiation magnetohydrodynamic
  simulation of seconds-long black hole-neutron star mergers},''
  \href{http://dx.doi.org/10.1103/PhysRevD.106.023008}{{\em \prd} {\bfseries
  106} no.~2, (July, 2022) 023008},
  \href{http://arxiv.org/abs/2111.04621}{{\ttfamily arXiv:2111.04621
  [astro-ph.HE]}}.

\bibitem{palenzuela15}
C.~{Palenzuela}, S.~L. {Liebling}, D.~{Neilsen}, L.~{Lehner}, O.~L.
  {Caballero}, E.~{O'Connor}, and M.~{Anderson}, ``{Effects of the
  microphysical equation of state in the mergers of magnetized neutron stars
  with neutrino cooling},''
  \href{http://dx.doi.org/10.1103/PhysRevD.92.044045}{{\em \prd} {\bfseries 92}
  no.~4, (Aug., 2015) 044045},
  \href{http://arxiv.org/abs/1505.01607}{{\ttfamily arXiv:1505.01607 [gr-qc]}}.

\bibitem{palenzuela18}
C.~{Palenzuela}, B.~{Mi{\~n}ano}, D.~{Vigan{\`o}}, A.~{Arbona},
  C.~{Bona-Casas}, A.~{Rigo}, M.~{Bezares}, C.~{Bona}, and J.~{Mass{\'o}}, ``{A
  Simflowny-based finite-difference code for high-performance computing in
  numerical relativity},''
  \href{http://dx.doi.org/10.1088/1361-6382/aad7f6}{{\em Classical and Quantum
  Gravity} {\bfseries 35} no.~18, (Sept., 2018) 185007},
  \href{http://arxiv.org/abs/1806.04182}{{\ttfamily arXiv:1806.04182
  [physics.comp-ph]}}.

\bibitem{Kastaun:2020uxr}
W.~Kastaun, J.~V. Kalinani, and R.~Ciolfi, ``{Robust Recovery of Primitive
  Variables in Relativistic Ideal Magnetohydrodynamics},''
  \href{http://dx.doi.org/10.1103/PhysRevD.103.023018}{{\em Phys. Rev. D}
  {\bfseries 103} no.~2, (2021) 023018},
  \href{http://arxiv.org/abs/2005.01821}{{\ttfamily arXiv:2005.01821 [gr-qc]}}.

\bibitem{PhysRevLett.97.141101}
L.~Baiotti and L.~Rezzolla, ``Challenging the paradigm of singularity excision
  in gravitational collapse,''
  \href{http://dx.doi.org/10.1103/PhysRevLett.97.141101}{{\em Phys. Rev. Lett.}
  {\bfseries 97} (Oct, 2006) 141101}.
  \url{https://link.aps.org/doi/10.1103/PhysRevLett.97.141101}.

\bibitem{Shibata_2008}
M.~Shibata and K.~Taniguchi, ``Merger of black hole and neutron star in general
  relativity: Tidal disruption, torus mass, and gravitational waves,''
  \href{http://dx.doi.org/10.1103/physrevd.77.084015}{{\em Physical Review D}
  {\bfseries 77} no.~8, (Apr, 2008) }.
  \url{https://doi.org/10.1103%2Fphysrevd.77.084015}.

\bibitem{Kyutoku_2015}
K.~Kyutoku, K.~Ioka, H.~Okawa, M.~Shibata, and K.~Taniguchi, ``Dynamical mass
  ejection from black hole-neutron star binaries,''
  \href{http://dx.doi.org/10.1103/physrevd.92.044028}{{\em Physical Review D}
  {\bfseries 92} no.~4, (Aug, 2015) }.
  \url{https://doi.org/10.1103%2Fphysrevd.92.044028}.

\bibitem{Hayashi:2022cdq}
K.~Hayashi, K.~Kiuchi, K.~Kyutoku, Y.~Sekiguchi, and M.~Shibata,
  ``{General-relativistic neutrino-radiation magnetohydrodynamics simulation of
  seconds-long black hole-neutron star mergers: Dependence on the initial
  magnetic field strength, configuration, and neutron-star equation of
  state},'' \href{http://dx.doi.org/10.1103/PhysRevD.107.123001}{{\em Phys.
  Rev. D} {\bfseries 107} no.~12, (2023) 123001},
  \href{http://arxiv.org/abs/2211.07158}{{\ttfamily arXiv:2211.07158
  [astro-ph.HE]}}.

\bibitem{arbona13}
A.~{Arbona}, A.~{Artigues}, C.~{Bona-Casas}, J.~{Mass{\'o}}, B.~{Mi{\~n}ano},
  A.~{Rigo}, M.~{Trias}, and C.~{Bona}, ``{Simflowny: A general-purpose
  platform for the management of physical models and simulation problems},''
  \href{http://dx.doi.org/10.1016/j.cpc.2013.04.012}{{\em Computer Physics
  Communications} {\bfseries 184} (Oct., 2013) 2321--2331}.

\bibitem{arbona18}
A.~{Arbona}, B.~{Mi{\~n}ano}, A.~{Rigo}, C.~{Bona}, C.~{Palenzuela},
  A.~{Artigues}, C.~{Bona-Casas}, and J.~{Mass{\'o}}, ``{Simflowny 2: An
  upgraded platform for scientific modelling and simulation},''
  \href{http://dx.doi.org/10.1016/j.cpc.2018.03.015}{{\em Computer Physics
  Communications} {\bfseries 229} (Aug., 2018) 170--181},
  \href{http://arxiv.org/abs/1702.04715}{{\ttfamily arXiv:1702.04715 [cs.MS]}}.

\bibitem{hornung02}
R.~D. Hornung and S.~R. Kohn, ``Managing application complexity in the samrai
  object-oriented framework,'' \href{http://dx.doi.org/10.1002/cpe.652}{{\em
  Concurrency and Computation: Practice and Experience} {\bfseries 14} no.~5,
  (2002) 347--368}. \url{http://dx.doi.org/10.1002/cpe.652}.

\bibitem{gunney16}
B.~T. Gunney and R.~W. Anderson, ``Advances in patch-based adaptive mesh
  refinement scalability,''
  \href{http://dx.doi.org/https://doi.org/10.1016/j.jpdc.2015.11.005}{{\em
  Journal of Parallel and Distributed Computing} {\bfseries 89} (2016) 65 --
  84}.
  \url{http://www.sciencedirect.com/science/article/pii/S0743731515002129}.

\bibitem{vigano19}
D.~{Vigan{\`o}}, D.~{Mart{\'\i}nez-G{\'o}mez}, J.~A. {Pons}, C.~{Palenzuela},
  F.~{Carrasco}, B.~{Mi{\~n}ano}, A.~{Arbona}, C.~{Bona}, and J.~{Mass{\'o}},
  ``{A Simflowny-based high-performance 3D code for the generalized induction
  equation},'' \href{http://dx.doi.org/10.1016/j.cpc.2018.11.022}{{\em Computer
  Physics Communications} {\bfseries 237} (Apr, 2019) 168--183},
  \href{http://arxiv.org/abs/1811.08198}{{\ttfamily arXiv:1811.08198
  [astro-ph.IM]}}.

\bibitem{shu98}
C.-W. Shu, {\em Essentially non-oscillatory and weighted essentially
  non-oscillatory schemes for hyperbolic conservation laws},
  \href{http://dx.doi.org/10.1007/BFb0096355}{pp.~325--432}.
\newblock Springer Berlin Heidelberg, Berlin, Heidelberg, 1998.
\newblock \url{https://doi.org/10.1007/BFb0096355}.

\bibitem{suresh97}
A.~Suresh and H.~Huynh, ``Accurate monotonicity-preserving schemes with
  runge–kutta time stepping,''
  \href{http://dx.doi.org/https://doi.org/10.1006/jcph.1997.5745}{{\em Journal
  of Computational Physics} {\bfseries 136} no.~1, (1997) 83 -- 99}.
  \url{http://www.sciencedirect.com/science/article/pii/S0021999197957454}.

\bibitem{McCorquodale:2011}
P.~McCorquodale and P.~Colella, ``A high-order finite-volume method for
  conservation laws on locally refined grids,''
  \href{http://dx.doi.org/10.2140/camcos.2011.6.1}{{\em Commun. Appl. Math.
  Comput. Sci.} {\bfseries 6} no.~1, (2011) 1--25}.
  \url{https://doi.org/10.2140/camcos.2011.6.1}.

\bibitem{Mongwane:2015}
B.~{Mongwane}, ``{Toward a consistent framework for high order mesh refinement
  schemes in numerical relativity},''
  \href{http://dx.doi.org/10.1007/s10714-015-1903-7}{{\em General Relativity
  and Gravitation} {\bfseries 47} (May, 2015) 60},
  \href{http://arxiv.org/abs/1504.07609}{{\ttfamily arXiv:1504.07609 [gr-qc]}}.

\bibitem{rezbish}
N.~T. {Bishop} and L.~{Rezzolla}, ``{Extraction of gravitational waves in
  numerical relativity},''
  \href{http://dx.doi.org/10.1007/s41114-016-0001-9}{{\em Living Reviews in
  Relativity} {\bfseries 19} (Oct., 2016) 2},
  \href{http://arxiv.org/abs/1606.02532}{{\ttfamily arXiv:1606.02532 [gr-qc]}}.

\bibitem{brugman}
B.~Br\"ugmann, J.~A. Gonz\'alez, M.~Hannam, S.~Husa, U.~Sperhake, and W.~Tichy,
  ``Calibration of moving puncture simulations,''
  \href{http://dx.doi.org/10.1103/PhysRevD.77.024027}{{\em Phys. Rev. D}
  {\bfseries 77} (Jan, 2008) 024027}.
  \url{https://link.aps.org/doi/10.1103/PhysRevD.77.024027}.

\bibitem{1968JETP...26.1031K}
A.~P. {Kazantsev}, ``{Enhancement of a Magnetic Field by a Conducting Fluid},''
  {\em Soviet Journal of Experimental and Theoretical Physics} {\bfseries 26}
  (May, 1968) 1031.

\bibitem{1992ApJ...396..606K}
R.~M. {Kulsrud} and S.~W. {Anderson}, ``{The Spectrum of Random Magnetic Fields
  in the Mean Field Dynamo Theory of the Galactic Magnetic Field},''
  \href{http://dx.doi.org/10.1086/171743}{{\em \apj} {\bfseries 396} (Sept.,
  1992) 606}.

\bibitem{paschalidis2015one}
V.~Paschalidis, W.~E. East, F.~Pretorius, and S.~L. Shapiro, ``One-arm spiral
  instability in hypermassive neutron stars formed by dynamical-capture binary
  neutron star mergers,'' {\em Physical Review D} {\bfseries 92} no.~12, (2015)
  121502.

\bibitem{2016PhRvD..94f4011R}
D.~{Radice}, S.~{Bernuzzi}, and C.~D. {Ott}, ``{One-armed spiral instability in
  neutron star mergers and its detectability in gravitational waves},''
  \href{http://dx.doi.org/10.1103/PhysRevD.94.064011}{{\em \prd} {\bfseries 94}
  no.~6, (Sept., 2016) 064011},
  \href{http://arxiv.org/abs/1603.05726}{{\ttfamily arXiv:1603.05726 [gr-qc]}}.

\bibitem{east2016relativistic}
W.~E. East, V.~Paschalidis, F.~Pretorius, and S.~L. Shapiro, ``Relativistic
  simulations of eccentric binary neutron star mergers: One-arm spiral
  instability and effects of neutron star spin,'' {\em Physical Review D}
  {\bfseries 93} no.~2, (2016) 024011.

\bibitem{2016PhRvD..94d3003L}
L.~{Lehner}, S.~L. {Liebling}, C.~{Palenzuela}, and P.~M. {Motl}, ``{m =1
  instability and gravitational wave signal in binary neutron star mergers},''
  \href{http://dx.doi.org/10.1103/PhysRevD.94.043003}{{\em \prd} {\bfseries 94}
  no.~4, (Aug., 2016) 043003},
  \href{http://arxiv.org/abs/1605.02369}{{\ttfamily arXiv:1605.02369 [gr-qc]}}.

\end{thebibliography}\endgroup

\end{document}